\documentclass[12pt,fleqn,abstract]{scrartcl}%
	\usepackage[T1]{fontenc}
	\usepackage[utf8]{inputenc}
	\usepackage[english]{babel}
	\usepackage[english=american]{csquotes}
	\usepackage[pdftex]{graphicx}
	\usepackage{float}
	\usepackage{amssymb,amsmath}
	\usepackage{fancyhdr}
	\usepackage{textcomp}
	\usepackage{color}
	\usepackage{upgreek}
	\usepackage{stackrel}
 	\usepackage[small,nohug,heads=LaTeX]{diagrams} 
	\usepackage[pdftex,plainpages=false,pdfborder={0 0 0},pdfpagelabels,unicode=true,breaklinks=true]{hyperref} 
	
	\usepackage[amsmath,thmmarks,thref,hyperref]{ntheorem}

\theoremstyle{plain}
\newtheorem{Theorem}{Theorem}[section]

\newtheorem{Lemma}[Theorem]{Lemma}

\theorembodyfont{\upshape}
\newtheorem{Remark}[Theorem]{Remark}
\newtheorem{Example}[Theorem]{Example}
\newtheorem{Definition}[Theorem]{Definition}

\theoremstyle{nonumberplain}
\theorembodyfont{\itshape}

\theoremheaderfont{\itshape}
\theorembodyfont{\upshape}
\theoremsymbol{\ensuremath{_\Box}}
\newtheorem{Proof}{Proof.}

\providecommand{\DD}{\mathbb{D}}

\providecommand{\NN}{\mathbb{N}}

\providecommand{\RR}{\mathbb{R}}

\newcommand{\cC}{\mathcal{C}}
\newcommand{\cD}{\mathcal{D}}

\newcommand{\cM}{\mathcal{M}}

\newcommand{\cO}{\mathcal{O}}

\newcommand{\cT}{\mathcal{T}}

\newcommand{\cW}{\mathcal{W}}

\newcommand{\cZ}{\mathcal{Z}}

\newcommand{\fo}{\mathfrak{o}}

\newcommand{\fs}{\mathfrak{s}}

\providecommand{\abs}[1]{\left \lvert #1 \right \rvert}
\providecommand{\sabs}[1]{\lvert #1 \vert}
\providecommand{\babs}[1]{\bigl \lvert #1 \bigr \rvert}

\providecommand{\norm}[1]{\left \lVert #1 \right \rVert}

\providecommand{\scpro}[2]{\left \langle #1 , #2 \right \rangle}

\DeclareMathOperator\im{im}
\DeclareMathOperator\spann{span}

\DeclareMathOperator\dist{dist}
\DeclareMathOperator\tr{tr}

\DeclareMathOperator\id{id}

\DeclareMathOperator\dive{div}
\DeclareMathOperator{\grad}{grad}

\providecommand{\dd}{\mathrm{d}}
\providecommand{\DD}{\mathrm{D}}
\providecommand{\e}{\mathrm{e}}

\providecommand{\ie}{i.e.\ }
\providecommand{\eg}{e.g.\ }
\providecommand{\cf}{cf.\ }

\SetMathAlphabet{\mathcal}{normal}{OMS}{cmsy}{m}{n}

\DeclareSymbolFont{lettersA}{U}{txmia}{m}{it}
\DeclareMathSymbol{\updelta}{\mathord}{lettersA}{"0E}

\DeclareMathOperator{\proj}{P}  
\DeclareMathOperator{\sff}{II}  
\DeclareMathOperator\rank{rank}  

\newcommand{\sH}{\mathsf{H}}
\newcommand{\sN}{\mathsf{N}}
\newcommand{\sT}{\mathsf{T}}
\newcommand{\sV}{\mathsf{V}}

\newcommand{\Heff}{H_\mathrm{eff}}

\newcommand{\D}{\mathrm{d}}

\newcommand{\projTB}{\proj^{\sT B}}
\newcommand{\projNB}{\proj^{\sN B}}

\newcommand{\projHM}{\proj^{\sH M}}
\newcommand{\projVrM}{\proj^{\sV {\mathring{M}}}}
\newcommand{\projHrM}{\proj^{\sH {\mathring{M}}}}

\newcommand{\Vb}{V_\mathrm{bend}}
\newcommand{\eps}{\varepsilon}

\makeatletter
\renewcommand*\env@matrix[1][*\c@MaxMatrixCols c]{%
  \hskip -\arraycolsep
  \let\@ifnextchar\new@ifnextchar
  \array{#1}}
\makeatother

\begin{document}

\title{Generalised Quantum Waveguides}
\author{Stefan Haag, Jonas Lampart, Stefan Teufel}

\maketitle
\thispagestyle{empty}

\vspace{-9mm}
\begin{center}
		  Eberhard Karls Universit\"at T\"ubingen, Mathematisches Institut, \linebreak
	Auf der Morgenstelle 10, 72076 T\"ubingen, Germany. \linebreak
		
\end{center}

\begin{abstract}
We study general quantum waveguides and establish explicit effective Hamiltonians for the Laplacian on these spaces. 
A conventional quantum waveguide is an $\eps$-tubular neighbourhood of a curve in $\RR^3$ and the object of interest is the Dirichlet Laplacian  {on} 
this tube in the asymptotic limit $\eps\to0$.
We generalise this by considering fibre bundles $M$ over a $d$-dimensional submanifold $B\subset\RR^{d+k}$ {with} fibres diffeomorphic to $F\subset\RR^k$, whose total space is embedded into an $\eps$-neighbourhood of $B$.
From this point of view $B$ takes the role of the curve and $F$ that of the disc-shaped cross-section of a conventional quantum waveguide. 
  Our approach allows, {among other things}, for waveguides whose cross-sections $F$ are deformed along $B$ and also the study of the Laplacian on the boundaries of such waveguides.
By applying recent results on the adiabatic limit of Schr\"odinger operators on fibre bundles we show, in particular,  that for small energies the dynamics and the spectrum of the Laplacian on $M$  are reflected by the adiabatic approximation associated to the ground state band of the normal Laplacian. We give   explicit formulas for the according effective operator on $L^2(B)$ in various scenarios, thereby  improving and extending many of the known  results on quantum waveguides and quantum layers in $\RR^3$.
\end{abstract}
\newpage 

\tableofcontents

\section{Introduction}

Quantum waveguides have been studied by physicists, chemists and mathematicians for many years now and the rate at which new contributions appear is still high 
(see~\cite{BMT07, dV11, DE95, KS12, KR13, SS13} and references therein).
Mathematically speaking, a conventional quantum waveguide corresponds to the study of the  Dirichlet Laplacian on a thin tube around a smooth curve in $\RR^3$. Of particular interest are effects of the geometry of the tube on the spectrum of and the unitary group generated by the Laplacian.
Similarly so-called quantum layers, \ie the Laplacian on a thin layer around a smooth surface, have been studied~\cite{CEK04, KL12, KRT13}. The related problem of the constraining of a quantum particle to a neighbourhood of such a curve (or surface)  by a steep potential rather than through the boundary condition was studied in~\cite{daC, deO13, FrHe,JK71, Mar, Mit, WaTe}. Recently, progress has also been made on quantum waveguides and layers in magnetic fields~\cite{deO13, KR13, KRT13}.

There are   obvious geometric generalisations of these concepts. One can consider the Dirichlet Laplacian on small neighbourhoods of $d$-dimensional submanifolds of $\RR^{d+k}$ (see e.g.~\cite{LL06}),  or of any $(d+k)$-dimensional Riemannian manifold. Another possibility is to look at the Laplacian on the boundary of such a submanifold, which, in the case of a conventional waveguide, is a cylindrical surface around a curve in $\RR^3$. Beyond generalising to higher dimension and codimension,  one could also ask for waveguides with cross-sections that change their shape and size along the curve, or more generally along the submanifold around which the waveguide is modelled. 

In the majority of mathematical works on quantum waveguides, with the exception of~\cite{dV11}, such variations of the cross-section along the curve must be excluded. The reason is that, physically speaking, localizing a quantum particle to a thin domain leads to large kinetic energies in the constrained directions, \ie in the directions  normal to the curve for a conventional waveguide, and that variations of the cross-section lead to exchange of this kinetic energy between normal and tangent directions.
However, the common approaches  require that  the Laplacian acts only on functions that have much smaller derivatives in the tangent directions than in the normal directions.

In this paper we show how to cope with several of the possible generalisations mentioned above: (i) We consider general dimension and codimension of the submanifold along which the waveguide is modelled. (ii) We allow for general variations of the cross-sections along the submanifold and thus necessarily for kinetic energies of the same order in all directions, with possible exchange of energy between the tangent and normal directions. (iii) We also include the case of \enquote{hollow} waveguides, \ie the Laplacian on the boundary of a general \enquote{massive} quantum waveguide.

All of this is achieved by developing a suitable geometric framework for general quantum waveguides and by the subsequent application of recent results on the adiabatic limit of Schr\"odinger operators on fibre bundles \cite{LT2014}. As concrete applications we will mostly emphasise geometric effects and explain, in particular, how the known effects of \enquote{bending} and \enquote{twisting} of waveguides in $\RR^3$ manifest themselves in higher dimensional generalised waveguides. 

Before going into a more detailed discussion of our results and of the vast literature, let us review the main concepts in the context of conventional quantum waveguides in a geometrical language that is already adapted to our subsequent generalisation. Moreover, it will allow us to explain the adiabatic structure of the problem within a simple example.

Consider a smooth  curve  $c:\RR\to \RR^3$  parametrised by arclength with bounded second derivative $c''$ and its $\eps$-neighbourhood 
\[
\cT^\eps:=\bigl\{y\in\RR^3:\ \dist(y,B)\leq \eps\bigr\}\subset\RR^3 
\]
for some $\eps>0$. By  $B:=c(\RR)$ we denote  the image of the curve in $\RR^3$ and we call $\cT^\eps$ the  {tube} of a conventional waveguide.
The aim is to understand the Laplace operator $\Delta_{\updelta^3}$ on $L^2(\cT^\eps,\D \updelta^3)$ with Dirichlet boundary conditions  on $\cT^\eps$  in the asymptotic limit $\eps\ll 1$. As different metrics will appear in the course of the discussion, we make the Euclidean metric   $\updelta^3$ explicit in the Laplacian.

For $\eps$ small enough one can map $\cT^\eps$ diffeomorphically onto the $\eps$-tube in the normal bundle of $B$. In order to make the following compuations explicit, we pick an orthonormal frame along the curve. A natural choice is to start with an orthonormal basis $(\tau, e_1,e_2)$ at one point in $B$ such that $\tau=c'$ is tangent and $(e_1,e_2)$ are normal to the curve. Then one obtains   a (in this special case global) unique frame  by parallel transport of $(\tau, e_1,e_2)$ along the curve $B$. This construction is sometimes called the
relatively parallel adapted frame  \cite{Bishop1975}. The  frame $(\tau(x),e_1(x),e_2(x))$  satisfies the   differential equation
\begin{equation}
\label{eq:DGLBishop}
\begin{pmatrix} \tau' \\ e'_1 \\ e'_2 \end{pmatrix} = \begin{pmatrix} 0 & \kappa^1 & \kappa^2 \\ - \kappa^1 & 0 & 0 \\ -\kappa^2 & 0 & 0 \end{pmatrix} \begin{pmatrix} \tau \\ e_1 \\ e_2 \end{pmatrix}
\end{equation}
with the components of the mean curvature vector  $\kappa^\alpha:B\to\RR$ ($\alpha=1,2$) given by
\[
\kappa^\alpha(x) := \scpro{\tau'(x)}{e_\alpha(x)}_{\RR^3} = \scpro{c''(x)}{e_\alpha(x)}_{\RR^3} \, .
\]
The two normal vector fields $e_{1,2}:B\to \RR^3$ form an orthonormal frame of $B$'s normal bundle $\sN B$. Hence, for $\eps>0$ small enough, there is a canonical identification of the $\eps$-tube in the normal bundle denoted by 
 \[
M^\eps := \left\{ \bigl(x, n^1e_1(x)+n^2e_2(x)\bigr)\in \sN B:\  (n_1)^2+(n_2)^2\leq\eps^2\right\} \subset \sN B
\]
 with the original $\eps$-tube $\cT^\eps\subset\RR^3$ via the map
\begin{equation}
\label{eq:Phiconv}
\Phi:M^\eps\to\cT^\eps\, ,\quad \Phi :\bigl(x,n^1 e_1(x)+n^2 e_2(x)\bigr)\mapsto x + n^1 e_1(x)+n^2 e_2(x)\, .
\end{equation}
We will refer to $F^\eps_x:= M^\eps\cap \sN_x B$ as the cross-section of $M^\eps$   and to $\Phi(F^\eps_x)$ as  the  cross-section of $\cT^\eps$ at $x\in B$.

In order to give somewhat more substance to the simple example, let us generalise the concept of a conventional waveguide already at this point. For a smooth   function $f:B\to [f_-,f_+]$ with $0<f_-<f_+<\infty$ let 
\[
M^\eps_f := \left\{ \bigl(x, n^1e_1(x)+n^2e_2(x)\bigr)\in \sN B:\ (n_1)^2+(n_2)^2\leq\eps^2 \,f(x)^2\right\}  
\]
be the tube with varying cross-section  $F^\eps_x$, a disc of radius $ \eps f(x)$. This gives rise to a corresponding tube $\cT^\eps_f := \Phi(M^\eps_f)$ in $\RR^3$.
To not overburden notation, we will drop the subscript $f$ in the following, \ie put $M^\eps := M^\eps_f$ and $\cT^\eps:=\cT^\eps_f $.

By equipping $M^\eps$ with the pullback metric  $g:=\Phi^*\updelta^3$, we can turn $\Phi$ into an isometry. Then the Dirichlet Laplacian   $\Delta _{\updelta^3}$     on $L^2(\cT^\eps,\D\updelta^3)$ is unitarily equivalent  to  the Dirichlet Laplacian   $\Delta _g$ on $L^2(M^\eps,\D g)$. 

In order to obtain an explicit expression for $\Delta _g$ with respect to the 
bundle coordinates $(x,n^1,n^2)$  associated with the orthonormal frame $(e_1(x), e_2(x))$,  
we need to compute the pullback metric $g$ on the tube $M^\eps$. For the coordinate vector fields $\partial_x$ and $\partial_{n^\alpha}$, $\alpha\in\{1,2\}$, one finds
\begin{align*}
\Phi_*\partial_x|_{(x,n)} &= \tfrac{\dd}{\dd x} \Phi\Bigl(\bigl(c(x),n^\alpha e_\alpha\bigl(c(x)\bigr)\Bigr) = \tfrac{\dd}{\dd x} \Bigl(c(x) + n^\alpha e_\alpha\bigl(c(x)\bigr) \Bigr) \\
&= \tau(x) - n^\alpha \kappa^\alpha(x) \tau(x) = \bigl(1 - n\cdot \kappa(x)\bigr) \tau(x)\, , \\
\Phi_*\partial_{n^\alpha}|_{(x,n)} &=\tfrac{\dd}{\dd n^\alpha} \Phi\Bigl(\bigl(c(x),n^\alpha e_\alpha\bigl(c(x)\bigr)\Bigr) = \tfrac{\dd}{\dd n^\alpha} \Bigl(c(x) + n^\alpha e_\alpha\bigl(c(x)\bigr) \Bigr) \\
&= e_\alpha(x)\,.
\end{align*}
Here, we used $c'(x)=\tau(x)$ and the differential equations~\eqref{eq:DGLBishop}. Knowing  that $(\tau,e_1,e_2)$ is an orthonormal frame of $\sT\RR^3|_B$ with respect to $\updelta^3$, this yields 
\[
g(x,n):=\begin{pmatrix} (1-n\cdot \kappa(x))^2 & 0 & 0 \\ 0 & 1 & 0 \\ 0 & 0 & 1 \end{pmatrix}.
\]
The Laplace-Beltrami operator on $M^\eps$ associated to $g$   is thus
\[
-\Delta _g = -\frac{1}{(1-n\cdot\kappa)^2}\left(\partial_x^2 + \frac{n\cdot\kappa'}{1-n\cdot\kappa}\,\partial_x\right) - \Delta_n + \frac{\kappa\cdot\nabla_n}{1-n\cdot\kappa}
\]
with $\Delta_n = \nabla_n^2 = \partial_{n^1}^2 + \partial_{n^2}^2$. 
As the Riemannian volume measure of $g$ on coordinate space reads $\D g  = (1-n\cdot\kappa(x)) \D \updelta^3$, it is convenient to introduce the multiplication operator $\cM_\rho:\psi\mapsto \rho^{1/2}\psi$ with the density $\rho(x,n):= 1-n\cdot\kappa(x)$ as a unitary operator from $L^2(M^\eps,\D g)$ to  $L^2(M^\eps,  \D\updelta^3)$. Here $\D\updelta^3$ just denotes Lebesgue measure on the coordinate space. A straightforward computation shows that
\begin{equation}
\label{eq:V_rho}
\cM_\rho\bigl(-\Delta _g\bigr) \cM_\rho^* = - \Delta_\mathrm{hor} - \Delta_n + V_\rho
\end{equation}
with
\[
\Delta_\mathrm{hor} := \partial_x\, \rho^{-2}\,\partial_x = \partial_x^2 + \partial_x (\rho^{-2} - 1) \partial_x
\]
and
\[
V_\rho := - \frac{\kappa^2}{4\rho^2} - \frac{n\cdot\kappa''}{2\rho^3} - \frac{5(n\cdot\kappa')^2}{4\rho^4}\,.
\]
The  rescaling by $\rho$ thus leads to a simpler split up of the Laplacian $-\Delta _g$ into a horizontal operator $\Delta_\mathrm{hor}$ and a vertical operator $\Delta_n$ given by  the Euclidean Laplace  operator  with Dirichlet boundary conditions on the cross-sections $F^\eps_x$. 
This simplification is, however,  at the expense 
 of an additional potential $V_\rho$.

Since $F^\eps_x$ is isometric to the disc of radius $\eps f(x)$, the  eigenvalues and eigenfunctions  of the vertical operator $\Delta_n$ are $\eps$-dependent and also functions of $x\in B$.
In order to arrive at an  $\eps$-independent vertical operator and an $\eps$-independent domain, one  dilates the fibres of $\sN B$ using $\cD_\eps:(x,n)\mapsto (x,\eps n)$  and its associated lift to a unitary operator mapping $L^2(M^{\eps=1},\D\updelta^3)$ to $L^2(M^\eps ,\D\updelta^3)$.
One then arrives at
\[
 \cD_\eps^*\cM_\rho \bigl(-\Delta _g\bigr) \cM_\rho^* \cD_\eps = - (\partial_x^2 + \eps   S^\eps) - \eps^{-2}\Delta_n + \Vb~
\]
with the second order differential operator 
\[
S^\eps := \eps^{-1}\partial_x \cD_\eps^* (\rho^{-2} - 1) \cD_\eps \partial_x  
= \eps^{-1}\partial_x \underbrace{(\rho_\eps^{-2} - 1)}_{= 2\eps n\cdot \kappa + \mathcal{O}(\eps^2)} \partial_x\, ,\quad \rho_\eps:= 1 - \eps n\cdot\kappa~\,
\]
and the bending potential
\begin{equation}
\label{eq:Vb intro}
\Vb := \cD_\eps^* V_\rho \cD_\eps = - \frac{\kappa^2}{4\rho_\eps^2} - \frac{\eps n\cdot\kappa''}{2\rho_\eps^3} - \frac{5(\eps n\cdot\kappa')^2}{4\rho_\eps^4} = - \frac{\kappa^2}{4} + \cO(\eps)\, .
\end{equation}
These terms account for the bending of the curve in the ambient space, i.e.\   its extrinsic geometry.  

In order to make   the asymptotic limit $\eps\to 0$ more transparent, we 
rescale units of energy in such a way that the transverse energies are of order one  by multiplying the full Laplacian with $\eps^2$. In summary one finds that $ -\eps^2\Delta _g  $ is unitarily equivalent to the operator
\[
H^\eps := \cD_\eps^* \cM_\rho\bigl(-\eps^2\Delta _g\bigr) \cM_\rho^* \cD_\eps = - \eps^2 \Delta_\sH - \Delta_\sV + \eps^2 \Vb - \eps^3 S^\eps\,
\]
acting on the domain $D(H^\eps) = W^2(M)\cap W^1_0(M)\subset L^2(M)$ with $M:=M^{\eps=1}$.
The Hamiltonian $H^\eps $ thus splits into the horizontal Laplacian $\eps^2\Delta_\sH :=\eps^2\partial_x^2$, the vertical Laplacian $\Delta_\sV :=\Delta_n$ and a additional differential operator $\eps H_1:=\eps^2 \Vb - \eps^3  S^\eps$ that will be treated as a perturbation.
This structure
is reminiscent of the starting point for the  Born-Oppenheimer approximation in molecular physics. There the $x$-coordinate(s) describe heavy nuclei and $\eps^2$ equals the inverse mass of the nuclei. The $n$-coordinates describe the electrons with mass of order one. 
In both cases the vertical resp.\ electron operator depends on $x$: the vertical Laplacian $\Delta_\sV $ in the quantum waveguide Hamiltonian $H^\eps$ depends on $x\in B$ through the domain $F_x:=F_x^{\eps=1}$ and the electron operator in the molecular Hamiltonian depends on $x$ through an interaction potential. 
This suggests to study the asymptotics $\eps\ll 1$ for quantum waveguide Hamiltonians by the 
same methods 
that have been successfully developed for molecular Hamiltonians, namely by adiabatic perturbation theory. The latter allows to seperate slow and fast degrees of freedom in a systematic way. In the context of quantum waveguides the tangent dynamics are slow compared to the frequencies of the  normal modes.

To illustrate the adiabatic structure of the problem, let $\lambda_0(x) \sim \frac{1}{f(x)^2}$ be the smallest eigenvalue 
of $-\Delta_\sV$ on $F_x$ and denote by 
$\phi_0(x)\in L^2(F_x)$ the corresponding normalised non-negative eigenfunction, the so-called ground state wave function. Let 
\[
P_0 L^2(M)  := \left\{ \Psi(x,n ) = \psi(x) \phi_0(x,n ):\ \psi\in L^2(B)\right\} \subset L^2(M)
\]
be the subspace of local product states and $P_0$ the orthogonal projection onto this space.
Now the restriction of $H^\eps$ to the subspace $P_0 L^2(M)$ is called the adiabatic approximation of $H^\eps$ on the ground state band and  the associated adiabatic operator is defined by  \begin{equation}\label{eq:H_a def}
 H_\mathrm{a}  := P_0H^\eps P_0\,.
\end{equation}
A simple computation using 
\[
(P_0\Psi) (x,n)= \langle \phi_0(x,\cdot),\Psi(x,\cdot)\rangle_{L^2(F_x)} \;\phi_0(x,n)
\]
and the unitary identification $W: P_0L^2(M) \to L^2(B)$, $ \psi(x) \phi_0(x,n^1,n^2)\mapsto \psi(x)$, shows that the adiabatic operator  $H_{\rm a}$ can be seen as an  operator acting only on functions on the curve~$B$, given by
\begin{align*}
(WH_{\rm a} W^*\psi )(x) &= \bigl( - \eps^2 \partial_x^2 + \lambda_0(x) + \eps^2 V_\mathrm{a}(x) + \eps^2   \Vb^0(x) \bigr) \psi(x)\\ 
& \hphantom{=}\ + \eps^3 \int_{F_x} \phi_0(x,n) \bigl(S^\eps  (\phi_0\psi)\bigr)(x,n)\ \D n+ \cO(\eps^3)\,,
\end{align*}
where $V_\mathrm{a}(x) := \norm{\partial_x \phi_0(x)}_{L^2(F_x)}^2$  and $\Vb^0(x)= - \frac{\kappa^2(x)}{4}$.
 As such it is a    one-dimensional Schr\"odinger-type operator with potential function $\lambda_0(x)+\mathcal{O}(\eps^2)$  and the asymptotic limit $\eps\ll 1$ corresponds to the  semi-classical limit. This analogy  shows that, in general, $- \eps^2 \partial_x^2 $
cannot be considered small compared to 
$\lambda_0(x) $, despite the factor of $\eps^2$. To see this, observe  that
 all  eigenfunctions $\psi^\eps$ (and also all  solutions of the corresponding time-dependent Schr\"odinger equation) are necessarily $\eps$-dependent with $ \|  \eps \partial_x  \psi^\eps\|^2\gg \eps^2$, unless $\lambda_0(x)\equiv c$ for some constant $c\in\RR$. To be more explicit, assume that $\lambda_0(x) \approx  {\omega^2}  (x-x_0)^2$ near a global minimum at $x_0$.
Then the lowest eigenvalues of $H_\mathrm{a}$ are  $e_\ell = \lambda_0(x_0) + \eps\omega (1 + 2 \ell) + \mathcal{O}(\eps^2)$ for $\ell=0,1,2,\dots$. While the level spacing of order $\eps$ is small compared to $\lambda_0$, it is large compared to the energy scale of order $\eps^2$ of the geometric potentials. And for states $\psi^\eps$ with $\ell \sim \eps^{-1}$ the kinetic energy  $ \|  \eps  \partial_x \psi^\eps\|^2 $ in the tangential direction is of order one.

However, the  majority of mathematical works on the subject considers   the situation where $ \|  \eps  \partial_x  \psi^\eps\|^2 $  is of order $\eps^2$. Clearly this only yields meaningful results if one assumes $\lambda_0 \equiv c$ for some constant $c\in\RR$. But this, in turn,  puts strong constraints on the possible geometries of the waveguide, which we avoid in the present paper.

Now the obvious mathematical question  is: {\em To what extent and in which sense do the properties of the   adiabatic operator $H_\mathrm{a}$ reflect the corresponding properties of $H^\eps$?} 
This question was answered in great generality in \cite{Lampart2013,LT2014} and we will translate these results to our setting of generalised quantum waveguides in Section~\ref{sec:mainresults}.  Roughly speaking, Theorem~\ref{thm:low spectrum} states  that the low-lying eigenvalues of $H_\mathrm{a}$ approximate those of $H^\eps$ up to errors of order $\eps^3$ in general, and up to order $\eps^4$ in the special case of $\lambda_0 \equiv c$. In the latter case the order $\eps^3$ terms in $H_\mathrm{a}$ turn out to be significant as well.

Our main  new contribution in this work  is to introduce the concept of  generalised quantum waveguides in Section~\ref{chap:general} and to compute explicitly the adiabatic operator 
for such generalised waveguides to all significant orders. For massive quantum waveguides, which are basically \enquote{tubes} with varying cross-sections modelled over submanifolds of arbitrary dimension and codimension, this is done in Section~\ref{chap:massiveQWG}. There we follow basically the same strategy as in the simple example given in the present section. We obtain general expressions for the adiabatic operator, from which we determine the relevant terms for different energy scales. Though the underlying calculations of geometric quantities have been long known~\cite{Tolar1988}, the contribution of $S^\eps$ has usually been neglected, because at the energy scale of $\eps^2$ it is of lower order than $\Vb$. This changes however on the natural energy scale of the example $M_f$ with non-constant $f$, where they may be of the same order, as we see in Section~\ref{sect:H_a alpha}. The contribution of the bending potential is known to be non-positive in dimensions $d=1$ or $d=2$~\cite{CEK04, Kre07}, while it has no definite 
sign in higher dimensions~\cite{Tolar1988}. It was stressed in~\cite{Kre07} that this leads to competing effects of bending and the non-negative \enquote{twisting potential} in quantum waveguides whose cross-sections $F_x$ are all isometric but not rotationally invariant and twist along the curve relative to the 
parallel frame. The generalisation of this twisting potential is the adiabatic potential $V_\mathrm{a}$, which is always non-negative and of the same order as $\Vb$. Using this general framework we generalise the concept of \enquote{twisted} waveguides to arbitrary dimension and codimension in Section~\ref{sect:twist}.

In Section~\ref{chap:hollowQWG} we finally consider hollow waveguides, which are   the boundaries of massive waveguides. So far there seem to be no results on these waveguides in the literature and the adiabatic operator derived in Section~\ref{sect:H_a hollow} is completely new. For hollow waveguides  the   vertical operator is essentially  the Laplacian on a compact manifold without boundary and thus its lowest eigenvalue vanishes identically, $\lambda_0(x) \equiv 0$. 
The adiabatic operator on $L^2(B)$  is quite different from the massive case.
Up to errors of order $\eps^3$, it is the sum of the Laplacian on $B$ and an effective potential given in~\eqref{eq:Vhollow}. For the special case of the boundary of $M^\eps_f$ discussed above this potential is given by 
\[
 \eps^2 \left[\tfrac12 \partial_x^2\log \bigl(2\uppi f(x)\bigr) + \tfrac14 \babs{\partial_x \log \bigl(2 \uppi f(x)\bigr)}^2 \right] 
=  \eps^2 \left[\tfrac12 \tfrac{f''}{f} - \tfrac14 \big(\tfrac{f'}{f}\big)^2\right]\,,
\]
which, in contrast to massive waveguides, is independent of the curvature $\kappa$ and depends only on the rate of change of Vol$(\partial F_x)= 2\uppi f(x)$. One can check for explicit examples that a local constriction in the tube, \eg for $f(x) = 2-\frac{1}{1+x^2}$, leads to an effective potential with wells. Thus, constrictions can support bound states on the surface of a tube.

\section{Generalised Quantum Waveguides}\label{chap:general}
In this part we give a precise definition of what we call generalised quantum waveguides.
In view of the example discussed in the introduction, the ambient space $\RR^3$ is replaced by $(d+k)$-dimensional Euclidean space and the role of the curve is played by an arbitrary smooth $d$-dimensional submanifold $B\subset\RR^{d+k}$. The generalised waveguide $M$ is contained in a neighbourhood of the zero section in $\sN B$ which can be diffeomorphically mapped to a tubular neighbourhood of $B\subset \RR^{d+k}$.
We will again call $F_x= M \cap \sN_x B$ the cross-section of the quantum waveguide at the point $x\in B$
and essentially assume that $F_x$ and $F_y$ are diffeomorphic for $x,y\in B$. This allows for general deformations of the cross-sections as one moves along the base, where in the introduction we only considered scaling by the function $f$.

In order to separate the Laplacian into its horizontal and vertical parts, we follow the strategy of the previous section. However, it will be more convenient to adopt the equivalent viewpoint, where we implement the scaling within the metric $g^\eps$ on   $M=M^{\eps=1}\subset \sN B$ instead of shrinking $M^\eps\subset \sN B$ and keeping $g $ fixed.
 
For the following considerations, we assume that there exists a tubular neighbourhood $B\subset \cT\subset\RR^{d+k}$ with globally fixed diameter, \ie there is $r>0$ such that normals to $B$ of length less than $r$ do not intersect. More precisely, we assume that the map
\[
\Phi:\sN B \to \RR^{d+k}\,,\quad (x,\nu)\mapsto x + \nu\,,
\]
restricted to
\[
\sN B^r :=  \bigl\{(x,\nu)\in\sN B:\ \norm {\nu}_{\RR^{d+k}} < r\bigr\}\subset\sN B
\]
is a diffeomorphism to its image $\cT$. Again, this mapping $\Phi$ provides a metric $G := \Phi^* \updelta^{d+k}$ on $\sN B^r$ and the rescaled version $G^\eps := \eps^{-2} \cD_\eps^* G$, where $\cD_\eps:(x,\nu)\mapsto(x,\eps \nu)$ is the dilatation of the fibres in $\sN B$.
\begin{Definition}
\label{def:HighDimQWG}
Let $B\subset\RR^{d+k}$ be a smooth $d$-dimensional submanifold with tubular neighbourhood $\cT\subset\RR^{d+k}$ and $F$ be a compact manifold with smooth boundary and  $\dim F\leq k$.

Suppose $M\subset\sN B^r=\Phi^{-1}(\cT)$ is a connected subset that is a fibre bundle with projection $\pi_M:M\to B$ and typical fibre $F$ such that the diagram
\begin{diagram}
 M    &\rInto    & \sN B   \\
 \dTo^{\pi_M}    &&    \dTo_{\pi_{\sN B}}\\
 B    &\rTo_{\id_B}&B
\end{diagram}
commutes. We then call the pair $(M,g^\eps)$, with the scaled pullback metric
\[
g^\eps:=G^\eps|_{\sT M}\in\cT^0_2(M)\,,
\]
a \emph{generalised quantum waveguide}.
\end{Definition}
It immediately follows from the commutative diagram that the cross-sections $F_x$ coincide with the fibres $\pi_M^{-1}(x)$ given by the fibre bundle structure. From now on we will usually refer to this object simply as the fibre of $M$ over $x$. Although other geometries are conceivable, the most interesting examples of generalised waveguides are given by subsets $M\subset \sN B^r$ of codimension zero and their boundaries.
In the following we will only treat these two cases and distinguish them by the following terminology:
\begin{Definition} \label{def:MassiveHollow}
Let $F\to M\xrightarrow{\pi_M}B$ be a generalised quantum waveguide as in Definition~\ref{def:HighDimQWG}. 
\begin{enumerate}
\item We call $M$ \emph{massive} if $F$ is the closure of an open, bounded and connected subset of $\RR^k$ with smooth boundary.
\item We call $M$ \emph{hollow} if $\mathrm{dim}(F)>0$ and there exists a massive quantum waveguide $\mathring{F}\to\mathring{M}\xrightarrow{\pi_{{\mathring{M}}}} B$ such that $M=\partial \mathring{M}$.
\end{enumerate}
\end{Definition}
This definition implies $\pi_M = \pi_{{\mathring{M}}}|_M$, \ie each fibre $F_x=\pi_M^{-1}(x)$ of a hollow quantum waveguide is the boundary of $\mathring{F}_x$, the fibre of the related massive waveguide $\mathring{M}$.

We denote by ${\sV M := \ker(\pi_{M*})\subset \sT M}$ the vertical subbundle of $\sT M$. Its elements are vectors that are tangent to the fibres of $M$. We refer to the orthogonal complement of $\sV M$ (with respect to $g:=g^{\eps=1}$) as the horizontal subbundle $\sH M \cong \pi^*_M(\sT B)$.
Clearly
\begin{equation}
\label{eq:THVM}
\sT M = \sH M \oplus \sV M\,,
\end{equation}
and this decomposition will turn out to be independent of $\eps$. That is, the decomposition $\sT M = \sH M \oplus \sV M$ is orthogonal for every $\eps>0$.
Furthermore, we will see (Lemma~\ref{lem:pullback} and equation~\eqref{eq:horblock} for the massive case, equation~\eqref{eq:g^eps hollow} for hollow waveguides) that the scaled pullback metric is always of the form
\begin{equation}
\label{eq:formg}
g^\eps = \eps^{-2}(\pi_M^* g_B + \eps h^\eps) +  g_F\,,
\end{equation}
where
\begin{itemize}
\item $g_B:= \updelta^{d+k}|_{\sT B}\in \cT^0_2(B)$ is the induced Riemannian metric on the submanifold $B$,
\item $h^\eps\in \cT^0_2(M)$ is a symmetric (but not necessarily non-degenerate) tensor with $h^\eps(V,\cdot)=0$ for any vertical vector field $V$,
\item $g_F:=g^\eps|_{\sV M}$ is the  $\eps$-independent   restriction of the scaled pullback metric to its vertical contribution.
\end{itemize}
Thus, if we define for any vector field $X\in\Gamma(\sT B)$ its unique horizontal lift $X^{\sH M}\in\Gamma(\sH M)$ by the relation  $\pi_{M*} X^{\sH M} = X$, we have
\begin{align*}
g^\eps(X^{\sH M},Y^{\sH M}) &= \eps^{-2}\bigl(g_B(X,Y)+\eps h^\eps(X^{\sH M},X^{\sH M})\bigr) \, , \\
g^\eps(X^{\sH M},V) &= 0 \, , \\
g^\eps(V,W) &= g_F(V,W)
\end{align*}
for all $X,Y\in\Gamma(\sT B)$ and $V,W\in \Gamma(\sV M)$.
\begin{Example} \label{ex:conv}
In the introduction we considered a massive waveguide $M=M_f$ with $d=1$ and $k=2$. The typical fibre was given by $F=\DD^2\subset\RR^2$. Using the bundle coordinates $(x,n^1,n^2)$ induced by~\eqref{eq:Phiconv}, one easily checks the the scaled pullback metric reads
\[
g^\eps:=\begin{pmatrix} \eps^{-2}(1-\eps n\cdot \kappa)^2 & 0 & 0 \\ 0 & 1 & 0 \\ 0 & 0 & 1 \end{pmatrix},
\]
hence 
\[
g_B=\dd x^2\, ,\quad g_F = \dd (n^1)^2 + \dd (n^2)^2\, ,\quad h^\eps = \big(-2 (n\cdot\kappa) + \eps (n\cdot\kappa)^2\big)\, \dd x^2\, .
\]
Finally remark that $\mathbb{S}^1=\partial \DD^2\subset\RR^2$ is the typical fibre of the associated hollow waveguide. 
\end{Example}
After having introduced the geometry of a generalised waveguide $(M,g^\eps)$, we now analyse the Laplace-Beltrami operator $\Delta_{g^\eps}$ with Dirichlet boundary conditions. The boundary condition is of course vacuous if $M$ is hollow since $\partial M = \emptyset$ in that case. Similarly to the introduction~\eqref{eq:V_rho}, we apply a unitary $\cM_{\rho_\eps}$ (which equals $\cD_\eps^* \cM_\rho \cD_\eps$ in the earlier notation). The according scaled density is given by
\begin{equation}
\label{eq:rhoepsgen}
\rho_\eps := {\frac{ {\D g^\eps}}{ {\D g^\eps_\mathrm{s}}}}\, ,\quad g^\eps_\mathrm{s} := \eps^{-2}\pi_M^* g_B + g_F\,.
\end{equation}
We call the metric $g_\mathrm{s}^\eps$ the (scaled) submersion metric on $M$, since it turns $\pi_M$ into a Riemannian submersion. The transformed Laplacian then reads 
\begin{align*}
H^\eps=\cM_{\rho_\eps}\bigl(-\Delta_{g^\eps}\bigr) \cM_{\rho_\eps}^* & = - \eps^2\Delta_\sH - \Delta_\sV + \eps^2 \Vb - \eps^3 S^\eps\,.
\end{align*}
Here, the horizontal Laplacian is defined by its quadratic form (with $g_\mathrm{s}:=g_\mathrm{s}^{\eps=1}$)
\begin{align*}
 \scpro{\Psi}{- \Delta_{\sH} \Psi} &= \int_M \pi^*_M g_B\bigl(\grad_{g_\mathrm{s}} \overline\Psi, \grad_{g_\mathrm{s}} \Psi\bigr)\ \dd g_\mathrm{s}\\
 &=\int_M g_\mathrm{s} \bigl(\grad_{g_\mathrm{s}} \overline\Psi, \projHM \grad_{g_\mathrm{s}} \Psi\bigr)\ \dd g_\mathrm{s}\, ,
\end{align*}
where $\projHM$ denotes the orthogonal projection to $\sH M$, so integration by parts yields (see also Section~\ref{sect:Laplace})
\begin{equation}
\label{eq:LaplaceH}
 \Delta_{\sH}= \dive_{g_\mathrm{s}} \projHM \grad_{g_\mathrm{s}}\, .
\end{equation}
The vertical operator is given on each fibre $F_x$ by the Laplace-Beltrami operator
\[
\Delta_\sV|_{F_x} := \Delta_{g_{F_x}}
\]
with Dirichlet boundary conditions. The bending potential
\begin{align}
\eps^2 \Vb &= \tfrac{1}{2} \dive_{g_\mathrm{s}^\eps} \grad_{g^\eps}(\log\rho_\eps) + \tfrac{1}{4} g^\eps(\dd \log\rho_\eps,\dd\log\rho_\eps) \nonumber \\
&= \tfrac{1}{2} (\eps^2\Delta_\sH + \Delta_\sV)(\log\rho_\eps)  + \tfrac{1}{4} g_F(\dd \log\rho_\eps,\dd\log\rho_\eps) + \cO(\eps^4) \label{eq:Vrhoeps}
\end{align}
is a by-product of the unitary transformation $M_{\rho_\eps}$ and the second order differential operator
\[
S^\eps: \Psi\mapsto S^\eps\Psi:= \eps^{-3}\dive_{g_\mathrm{s}}(g^\eps - g^\eps_\mathrm{s})(\dd \Psi,\cdot)
\]
accounts for the corrections to $g_s^\eps$.
\section{Adiabatic Perturbation Theory}\label{sec:mainresults}
In this section we show that the
adiabatic operator  $H_{\rm a}$ approximates essential features of generalised quantum waveguide Hamiltonians $H^\eps$, such as 
its unitary group and its spectrum.
This motivates the derivation of
explicit expansions of   $H_{\rm a}$ in the subsequent sections. 
In this work we will only consider the ground state band $\lambda_0(x)$ and pay special attention to the behaviour of $H^\eps$ for small energies.
This, as we will show, allows to view $H_{\rm a}$ as an operator on $L^2(B)$.
The results of this section were derived in~\cite{Lampart2013,LT2014} in more generality.

For a massive quantum waveguide set 
\begin{equation}
\label{eq:X1}
\tag{massive}
 H_F:=-\Delta_\sV
\end{equation}
and for a hollow waveguide 
\begin{equation}
\label{eq:X2}
\tag{hollow}
 H_F:=-\Delta_\sV + \tfrac{1}{2} \Delta_\sV (\log \rho_\eps) + \tfrac{1}{4} g_F(\dd \log\rho_\eps,\dd\log\rho_\eps)\, .
\end{equation}
Let $\lambda_0(x):=\min \sigma(H_{F_x})$ be the smallest eigenvalue of the fibre operator $H_F$ acting on the fibre over $x$. For hollow waveguides we have no boundary and $\lambda_0\equiv 0$ with the eigenfunction 
\[
\phi_0=\sqrt{\rho_\eps} \Big(\int_{F_x} \rho_\eps\ \dd g_F\Big)^{-1/2}= \pi_M^*\mathrm{Vol}(F_x)^{-1/2} + \mathcal{O}(\eps)\, .
\]
In the massive case we have $\lambda_0>0$ and denote by $\phi_0(x,\cdot)$ the uniquely determined positive normalised eigenfunction of $H_{F_x}$ with eigenvalue $\lambda_0(x)$. Let $P_0$ be the orthogonal projection in $L^2(M)$ defined by 
\[
(P_0 \Psi) ( x,\nu)=\phi_0\bigl(x,\nu\bigr) \int_{F_{x}}   \phi_0\bigl(x,\cdot\bigr) \Psi (x,\cdot\bigr)\ \dd g_F\,.
\]
The image of this projection is the subspace $L^2(B)\otimes \mathrm{span}(\phi_0) \cong L^2(B)$ of $L^2(M)$.
The function $\phi_0$ and its derivatives, both horizontal and vertical, are uniformly bounded in $\eps$. Thus, the action of the horizontal Laplacian $-\eps^2\Delta_\sH$ on $\phi_0$ gives a term of order $\eps$ and 
\begin{equation}
[H^\eps, P_0]P_0 = [H^\eps - H_F, P_0]P_0=\mathcal{O}(\eps)\label{eq:P0comm}
\end{equation}
as an operator from $D(H^\eps)$ to $L^2(M)$. Since this expression equals $(H^\eps-H_{\rm{a}})P_0$, this justifies the adiabatic approximation~\eqref{eq:H_a def} for states in the image of $P_0$. However, the error is of order $\eps$, while interesting effects of the geometry, such as the potentials $V_\mathrm{a}$ and $V_\mathrm{bend}$ discussed in the introduction, are of order $\eps^2$.
Because of this it is desirable to construct also a super-adiabatic approximation, consisting of a modified projection $P_\eps=P_0 + \mathcal{O}(\eps)\in \mathcal{L}(L^2(M))\cap\mathcal{L}(D(H^\eps))$ and an intertwining unitary $U_\eps$ with $P_\eps U_\eps  = U_\eps P_0$, such that the effective operator
\[
\Heff:=P_0 U_\eps^* H^\eps U_\eps P_0
\]
provides a better approximation of $H^\eps$ than $H_\mathrm{a}$ does. It then turns out that the approximation provided by $H_{\rm{a}}$ can also be made more accurate than expected from~\eqref{eq:P0comm} using the unitary $U_\eps$.

Such approximations can be constructed and justified if the geometry of $(M,g^\eps)$ satisfies some uniformity conditions. Here we only spell out the conditions relevant to our case, for a comprehensive discussion see~\cite{Lampart2013}.
\begin{Definition}
The generalised quantum waveguide $(M,g^\eps)$ is a \emph{waveguide of bounded geometry} if the following conditions are satisfied:
\begin{enumerate}
 \item The manifold $(B,g_B)$ is of bounded geometry. This means it has positive injectivity radius and for every $k\in \NN$ there exists a constant $C_k>0$ such that
\[
g_B(\nabla^k R, \nabla^k R)\leq C_k\,,
\]
where $R$ denotes the curvature tensor of $B$ and $\nabla$, $g_B$ are the connections and metrics induced on the tensor bundles over $B$.
\item The fibre bundle $(M,g)\xrightarrow{\pi_M} (B, g_B)$ is uniformly locally trivial. That is, there exists a Riemannian metric $g_0$ on $F$ such that for every $x\in B$ and metric ball $B(r,x)$ of radius $r< r_\mathrm{inj}(B)$ there is a trivialisation $\Omega_{x,r}:(\pi_M^{-1}(B(x,r)), g)\to (B(x,r) \times F, g_B\times g_0)$, and the tensors $\Omega_{x,r}^*$ and $\Omega_{x,r*}$ and all their covariant derivatives are bounded uniformly in $x$.
\item The embeddings $(M,g) \hookrightarrow (\sN B,G)$ and $(B,g_B)\hookrightarrow \RR^{d+k}$ are bounded with all their derivatives.
\end{enumerate}
\end{Definition}
These conditions are trivially satisfied for compact manifolds $M$ and many examples such as \enquote{asymptotically straight} or periodic waveguides. The existence result for the super-adiabatic approximation can be formulated as follows.
\begin{Theorem}[\cite{LT2014}]
Let $M$ be a waveguide of bounded geometry and set $\Lambda := \inf_{x\in B} \min (\sigma(H_{F_x})\setminus\lambda_0)$. For every $N\in \NN$ there exist a projection $P_\eps$ and a unitary $U_\eps$ in $\mathcal{L}(L^2(M))\cap\mathcal{L}(D(H^\eps))$, intertwining $P_0$ and $P_\eps$, such that 
for every $\chi\in \mathcal{C}^\infty_0\big((-\infty, \Lambda), [0,1])$, satisfying $\chi^p\in \cC^\infty_0\big((-\infty, \Lambda), [0,1])$ for every $p\in(0,\infty)$,  we have
\[
\norm{H^\eps \chi(H^\eps) - U_\eps \Heff \chi(\Heff) U_\eps^*} =\mathcal{O}(\eps^N)\,.
\]
In particular the Hausdorff distance between the spectra of $H^\eps$ and $\Heff$  is small,  i.e.\ for every $\delta>0:$ 
\[
 \dist\big(\sigma(H^\eps)\cap (-\infty, \Lambda-\delta],\sigma(\Heff)\cap (-\infty, \Lambda-\delta]\big)=\mathcal{O}(\eps^N)\,.
\]
\end{Theorem}
For $N=1$ we can choose $P_\eps=P_0$, so at first sight the approximation of $H^\eps$ by $H_\mathrm{a}$ yields errors of order $\eps$. More careful inspection shows that for $N>1$ we have $H_\mathrm{a}-\Heff=\mathcal{O}(\eps^2)$ as an operator from $W^2(B)$ to $L^2(M)$, so the statement on the spectrum holds for $H_\mathrm{a}$ with an error of order $\eps^2$. This improvement over~\eqref{eq:P0comm} relies on the existence of $U_\eps$ for a better choice of trial states.
Close to the ground state the approximation is even more accurate.
\begin{Theorem}[\cite{LT2014}]\label{thm:low spectrum}
Let $M$ be a waveguide of bounded geometry, $\Lambda_0:= \inf_{x\in B} \lambda_0(x)$ and ${0< \alpha\leq 2}$. Then for every $C>0$
\[
\dist\big(\sigma(H^\eps)\cap (-\infty, \Lambda_0 + C\eps^\alpha],\sigma(H_\mathrm{a})\cap (-\infty,  \Lambda_0 + C\eps^\alpha]\big)=\mathcal{O}(\eps^{2+\alpha/2})\,.
\]
Assume, in addition, that  $\Lambda_0 +C\eps^\alpha$ is strictly below the essential spectrum of 
$H_{\rm a}$ in the sense that for some $\delta>0$ and $\eps$ small enough the spectral projection ${\bf 1}_{(-\infty, \Lambda_0 + (C+\delta)\eps^\alpha]}(H_{\rm a})$ has finite rank.
Then, if  $\mu_0< \mu_1\leq \ldots \leq \mu_K$ are all the eigenvalues of $H_{\rm a}$ below $\Lambda_0+C\eps^\alpha$, $H^\eps$ has at least $K+1$ eigenvalues $\nu_0 <\nu_1 \leq \ldots \leq \nu_{K} $ below its essential spectrum and 
\[
|\mu_j-\nu_j| =\mathcal{O}(\eps^{2+\alpha}) 
\]
for $j\in \{0,\dots, K\}$.
\end{Theorem}
The natural energy scale $\alpha$ to consider this theorem would be the spacing of eigenvalues of $H_\mathrm{a}$. This of course depends on the specific situation. If $\lambda_0$ is constant we will see that $\alpha=2$ is a natural choice. In the somewhat more generic case in which the eigenband $\lambda_0(x)$ has a global and non-degenerate minimum, as in the example of the waveguide $M_f$ in the introduction, the lowest eigenvalues of $H_\mathrm{a}$ will behave like those of an harmonic oscillator and $\alpha=1$ is the correct choice of scale. In this case the set $(-\infty, \Lambda_0 + C \eps^2]\cap \sigma(H_\mathrm{a})$ will just be empty for $\eps$ small enough and thus by the theorem there is no spectrum of $H^\eps$ in this interval. 

We remark that results can be obtained also for energies higher than $\Lambda$ and projections to other eigenbands than $\lambda_0$. The relevant condition is that they are separated from the rest of the spectrum of $H_F$ by a local gap. For $\lambda_0$ this is a consequence of the bounded geometry of $M$ (see~\cite[Proposition 4.1]{LT2014}). The approximation of spectra is not mutual as for low energies, but there is always spectrum of $H^\eps$ near that of $\Heff$ (see~\cite[Corollary 2.4]{LT2014}). 

From now on we will focus on analysing the adiabatic operator. In particular we will see how the geometry of the waveguide enters into this operator and its expansion up to order $\eps^4$, which is relevant for small energies irrespective of the super-adiabatic corrections by Theorem~\ref{thm:low spectrum}. We now give a general expression for $H_\mathrm{a}$ from which we will derive the explicit form for various specific situations. First group the terms of $H^\eps$ in such a way that
\[
 H^\eps= -\eps^2 \Delta_\sH + H_F + \eps H_1\, ,
\]
by taking
\begin{align*}
\tag{massive} H_1&= - \eps^2 S^\eps + \eps \Vb\, , \\
\tag{hollow} H_1&
\begin{aligned}[t]
&= -\eps^2 S^\eps + \eps \Vb - \eps^{-1}\bigl(\tfrac{1}{2} \Delta_\sV (\log \rho_\eps) + \tfrac{1}{4} g_F(\dd \log\rho_\eps,\dd\log\rho_\eps)\big)\\
&= -\eps^2 S^\eps + \underbrace{\tfrac{\eps}{2} \Delta_\sH (\log \rho_\eps) + \mathcal{O}(\eps^3)}_{=:\eps\tilde V_{\rm{bend}}}\, .
\end{aligned}
\end{align*}
Projecting this expression with $P_0$ as in equation~\eqref{eq:H_a def} gives $H_FP_0 =\lambda_0 P_0$ and 
\begin{equation}
\label{eq:Va}
P_0\Delta_\sH P_0 = \Delta_{g_B} + \underbrace{\tfrac{1}{2} \mathrm{tr}_{g_B}(\nabla^B \bar\eta) -\int_{F_x} \pi_M^*g_B(\grad_{g_\mathrm{s}}\phi_0, \grad_{g_\mathrm{s}}\phi_0)\ \dd g_F}_{=: -V_\mathrm{a}}\, ,
\end{equation}
where $\nabla^B$ is the Levi-Civit\`{a} connection of $g_B$, $\bar \eta$ is the 
one-form
\[
 \bar\eta(X):=\int_{F_x} \vert \phi_0 \vert^2 g_B(\pi_{M*}\eta_F, X)\ \dd g_F
\]
and $\eta_F$ is the mean curvature vector of the fibres (see equation~\eqref{eq:etaF}). The derivation for the projection of $\Delta_\sH$ can be found in~\cite[Chapter 3]{Lampart2013}. 

Altogether we have the expression
\begin{equation}\label{eq:Ha}
 H_\mathrm{a}= -\eps^2\Delta_{g_B} + \lambda_0 + \eps^2 V_\mathrm{a} + \eps P_0 H_1 P_0
\end{equation}
for the adiabatic operator as an operator on $L^2(B)$. By analogy with the introduction, we view
\[
(P_0 S^\eps P_0)\psi= \int_{F_x} \phi_0  S^\eps (\phi_0\psi) \ \dd g_F
\]
as an operator on $B$ via the identification $L^2(B)\cong L^2(B) \otimes \mathrm{span}(\phi_0)$. By the same procedure, projecting the potentials in $H_1$ amounts to averaging them over the fibres with the weight $\abs{\phi_0}^2$.
\section{Massive Quantum Waveguides}\label{chap:massiveQWG}
The vast literature on quantum waveguides is, in our terminology, concerned with the case of massive waveguides. In this section we give a detailed derivation of the effects due to the extrinsic geometry of $B\subset \RR^{d+k}$. The necessary calculations of the metric $g^\eps$ and the bending potential $\Vb$ have been performed in all of the works on quantum waveguides for the respective special cases, and by Tolar~\cite{Tolar1988} for the leading order $\Vb^0$ in the general case. A generalisation to tubes in Riemannian manifolds is due to Wittich~\cite{Wit}.

We then discuss the explicit form of the adiabatic Hamiltonian~\eqref{eq:H_a def}, calculating the adiabatic potential and the projection of $H_1$. In particular we generalise the concept of a \enquote{twisted} waveguide (c.f.~\cite{Kre07}) to arbitrary dimension and codimension in Section~\ref{sect:twist}. We also examine the role of the differential operator $S^\eps$ in $H_\mathrm{a}$, which is rarely discussed in the literature, and its relevance for the different energy scales $\eps^\alpha$.
\subsection{The Pullback Metric}\label{sec:pullbackmassive}
Let $(x^1,\dots,x^d)$ be local coordinates on $B$ and $\{e_\alpha\}_{\alpha=1}^k$  a local orthonormal frame of $M$ with respect to $g_B^\bot:=\updelta^{d+k}|_{\sN B}$ such that every normal vector $\nu(x)\in \sN_x B$ may be written as
\begin{equation}
\label{eq:xn}
\nu(x) = n^\alpha e_\alpha(x)\, .
\end{equation}
These bundle coordinates yield local coordinate vector fields
\begin{equation}
\label{eq:productbasis}
\partial_i|_{(x,n)} := \tfrac{\partial}{\partial x^i}\,,\quad
\partial_{d+\alpha}|_{(x,n)} := \tfrac{\partial}{\partial n^\alpha}
\end{equation}
on $M$ for $i\in\{1,\dots,d\}$ and $\alpha \in \{1,\dots,k\}$. The aim is to obtain formulas for the coefficients of the unscaled pullback metric $g=\Phi^*\updelta^{d+k}|_{\sT M}$ with respect to these coordinate vector fields.

Let $I\subset\RR$ be an open neighbourhood of  zero, $b:I\to M$, $s\mapsto b(s) = (c(s),v(s))$ be a curve with $b(0)=(x,n)$ and $b'(0)=\xi\in \sT_{(x,n)}M$. It then holds that
\[
\Phi_*\xi = \left.\tfrac{\dd}{\dd s}\right|_{s=0} \Phi\bigl(b(s)\bigr) = c'(0) + v'(0)\, .
\]
For the case $\xi=\partial_i|_{(x,n)}$, we choose the curve $b:I\to M$ given by
\[
b(s) = \Bigl(c(s),n^\alpha e_\alpha\bigl(c(s)\bigr)\Bigr)\quad\Rightarrow\quad \Phi\bigl(b(s)\bigr) = c(s) + n^\alpha e_\alpha\bigl(c(s)\bigr)
\]
where $c:I\to B$ is a smooth curve with $c(0)=x$ and $c'(0) = \partial_{x^i}\in \sT_xB$. We then have
\[
\Phi_*\partial_i|_{(x,n)} = c'(0) + n^\alpha \left.\tfrac{\dd}{\dd s}\right|_{s=0} e_\alpha\bigl(c(s)\bigr) = \partial_{x^i} + n^\alpha \nabla^{\RR^{d+k}}_{\partial_{x^i}} e_\alpha(x)
\]
In order to relate the appearing derivative $\nabla^{\RR^{d+k}}_{\partial_{x^i}} e_\alpha(x)$ to the extrinsic curvature of $B$, we project the latter onto its tangent and normal component, respectively. Therefor, we introduce the Weingarten map
\[
\cW:\Gamma(\sN B)\to \cT^1_1(B)\, ,\quad e_\alpha \mapsto \cW(e_\alpha)\partial_{x^i}:=-\projTB \nabla^{\RR^{d+k}}_{\partial_{x^i}} e_\alpha\, ,
\]
and the $\fs\fo(k)$-valued local connection one-form associated to the normal connection $\nabla^\sN$ with respect to $\{e_\alpha\}_{k=1}^\alpha$, \ie
\[
\omega^\sN(\partial_{x^i})e_\alpha = \nabla^\sN_{\partial_{x^i}} e_\alpha := \projNB \nabla^{\RR^{d+k}}_{\partial_{x^i}} e_\alpha\, .
\]
With these objects we   have
\begin{equation}
 \label{eq:partiali}
\Phi_*\partial_i|_{(x,n)} =  \partial_{x^i} + n^\alpha \Bigl(- \cW\bigl(e_\alpha(x)\bigr)\partial_{x^i} + \omega^\sN(\partial_{x^i})e_\alpha(x)\Bigr).
\end{equation}
For the case $\xi=\partial_{d+\alpha}|_{(x,n)}$, one takes the curve $b:I\to M$ with
\[
b(s) = \bigl(x,\nu(x) + s e_\alpha(x)\bigr)\quad\Rightarrow\quad \Phi\bigl(b(s)\bigr) = x + (n^\beta + s \updelta_\alpha^{\hphantom{\alpha}\beta}) e_\beta(x)\, .
\]
Hence,
\begin{equation}
\label{eq:isoalpha}
\Phi_*\partial_{d+\alpha}|_{(x,n)} = 0 +  \left.\tfrac{\dd}{\dd s}\right|_{s=0} (n^\beta + s \updelta_\alpha^{\hphantom{\alpha}\beta}) e_\beta(x) = \updelta_\alpha^{\hphantom{\alpha}\beta} e_\beta(x) = e_\alpha(x)\, . 
\end{equation}
Combining the expressions for the tangent maps, we finally obtain the following expressions for the pullback metric $g$:

\begin{Lemma} \label{lem:pullback}
Let $(x,n)$ denote the local bundle coordinates on $M$ introduced in~\eqref{eq:xn} and~\eqref{eq:productbasis} the associated coordinate vector fields. Then the coefficients of the pullback metric are given by
\begin{align*}
g_{ij}(x,n) &= g_B(\partial_{x^i},\partial_{x^j}) - 2\sff(\nu)(\partial_{x^i},\partial_{x^j}) + g_B\bigl(\cW(\nu)\partial_{x^i},\cW(\nu)\partial_{x^j}\bigr) \\
&\hphantom{=}\,\,+g_B^\bot\bigl(\omega^\sN(\partial_{x^i})\nu,\omega^\sN(\partial_{x^j})\nu\bigr) \, , \\
g_{i,d+\alpha}(x,n) &= g_B^\bot\bigl(\omega^\sN(\partial_{x^i})\nu,e_\alpha\bigr)\, , \\
g_{d+\alpha,d+\beta}(x,n) &= g_B^\bot\bigl(e_\alpha,e_\beta\bigr) = \updelta_{\alpha\beta}
\end{align*}
for $i,j\in\{1,\dots,d\}$ and $\alpha,\beta\in\{1,\dots,k\}$. Here, ${\sff:\Gamma(\sN B)\to \cT^0_2(B)}$ stands for the second fundamental form defined by $\sff(\nu)(\partial_{x^i},\partial_{x^j}):=g_B(\cW(\nu)\partial_{x^i},\partial_{x^j})$.
\end{Lemma}
Let us now consider the scaled pullback metric $g^\eps=\eps^{-2}\cD_\eps^*\Phi^*\updelta^{d+k}|_{\sT M}$. Observe that for any $\xi\in \sT_{(x,n)}M$ 
\begin{align*}
\Phi_*(\cD_\eps)_*\xi &= (\Phi\circ \cD_\eps)_*\xi \\
&= \left.\tfrac{\dd}{\dd s}\right|_{s=0} (\Phi\circ \cD_\eps)\bigl(b(s)\bigr) 
= \left.\tfrac{\dd}{\dd s}\right|_{s=0} (\Phi\circ \cD_\eps)\bigl(c(s),v(s)\bigr) \\
&=\left.\tfrac{\dd}{\dd s}\right|_{s=0} \Phi\bigl(c(s),\eps v(s)\bigr) = c'(0) + \eps v'(0)
\end{align*}
and one immediately concludes from~\eqref{eq:partiali} and~\eqref{eq:isoalpha} that
\begin{align*}
\Phi_*(\cD_\eps)_*\partial_i|_{(x,n)} &= \partial_{x^i} + \eps n_\alpha \Bigl(- \cW\bigl(e_\alpha(x)\bigr)\partial_{x^i} + \omega^N(\partial_{x^i})e_\alpha(x)\Bigr), \\
\Phi_*(\cD_\eps)_*\partial_\alpha|_{(x,n)} &= \eps e_\alpha(x)\, .
\end{align*}
Consequently, the coefficients of the scaled pullback metric are given by
\begin{align*}
g_{ij}^\eps(x,n) &= \eps^{-2} \Bigl[g_B(\partial_{x^i},\partial_{x^j}) - \eps 2 \sff(\nu)(\partial_{x^i},\partial_{x^j}) + \eps^2 g_B\bigl(\cW(\nu)\partial_{x^i},\cW(\nu)\partial_{x^j}\bigr) \\
&\hphantom{=\frac{1}{\eps^2} \Bigl(} + \eps^2 g_B^\bot\bigl(\omega^\sN(\partial_{x^i})\nu,\omega^\sN(\partial_{x^j})\nu\bigr)\Bigr]\, , \\
g_{i,d+\alpha}^\eps(x,n) &= g_B^\bot\bigl(\omega^\sN(\partial_{x^i})\nu,e_\alpha\bigr)\, , \\
g_{d+\alpha,d+\beta}^\eps(x,n) &= \updelta_{\alpha\beta}
\end{align*}
for $i,j\in\{1,\dots,d\}$ and $\alpha,\beta\in\{1,\dots,k\}$.

We see that $\spann\{\partial_i|_{(x,n)}\}_{i=1}^d$ is not orthogonal to $\spann\{\partial_{d+\alpha}|_{(x,n)}\}_{\alpha=1}^k=\sV_{(x,n)}M$ with respect to $g^\eps$. However, any vector $\partial_i|_{(x,n)}$ can be orthogonalised by subtracting its vertical component. The resulting vector
\begin{align}
\partial_{x^i}^{\sH M}|_{(x,n)} &:= \partial_i|_{(x,n)} - g_{i,d+\beta}^\eps(x,n)\, {\partial_{d+\beta}|}_{(x,n)} \nonumber \\
&\hphantom{:}=  \partial_i|_{(x,n)} - g_B^\bot\bigl(\omega^\sN(\partial_{x^i})\nu,e_\beta\bigr)\, {\partial_{d+\beta}|}_{(x,n)} \label{eq:partialHM} \\
&\hphantom{:}= \partial_i|_{(x,n)} - g_{i,d+\beta}(x,n)\, {\partial_{d+\beta}|}_{(x,n)} \nonumber
\end{align}
is the horizontal lift of $\partial_{x^i}$. Consequently, the orthogonal complement of $\sV_{(x,n)}M$ with respect to $g^\eps$ is given by $\sH_{(x,n)}M=\spann\{\partial_{x^i}^{\sH M}|_{(x,n)}\}_{i=1}^d$ for all $\eps>0$.
Finally, a short computation shows that
\begin{align}
& g^\eps(\partial_{x^i}^{\sH M}|_{(x,n)},\partial_{x^j}^{\sH M}|_{(x,n)}) \nonumber \\
&\ =\eps^{-2} g_B\Bigl(\bigl(1-\eps\cW(\nu)\bigr)\partial_{x^i}, \bigl(1-\eps\cW(\nu)\bigr)\partial_{x^j}\Bigr) \label{eq:horblock} \\
&\ =\eps^{-2}\left[g_B(\partial_{x^i},\partial_{x^j}) +  \eps \Bigl(-2\sff(\nu)(\partial_{x^i},\partial_{x^j}) + \eps g_B\bigl(\cW(\nu)\partial_{x^i},\cW(\nu)\partial_{x^j}\bigr)\Bigr)\right]. \nonumber
\end{align}
Hence, the scaled pullback metric $g^\eps$ actually has the form~\eqref{eq:formg} with \enquote{horizontal correction}
\begin{equation}
\label{eq:heps}
h^\eps(\partial_{x^i}^{\sH M}|_{(x,n)},\partial_{x^j}^{\sH M}|_{(x,n)}) = - 2 \sff(\nu)(\partial_{x^i},\partial_{x^j}) + \eps g_B\bigl(\cW(\nu)\partial_{x^i},\cW(\nu)\partial_{x^j}\bigr)\, .
\end{equation}

\begin{Remark} \label{rem:complgeod}
The fibres $F_x$ of $M$ are completely geodesic for the pullback metric $g^\eps$. In order to see this, we show that the second fundamental form of the fibres $\sff^F|_x:\sH_x M \to \cT^0_2(F_x)$ vanishes identically. Since the latter is a symmetric tensor, it is sufficient to show that the diagonal elements
\[
\sff^F(\partial_{x^i}^{\sH M})(\partial_\alpha,\partial_\alpha) = g^\eps(\nabla^M_{\partial_\alpha} \partial_\alpha , \partial_{x^i}^{\sH M}) 
\]
are zero. 
Using Koszul's formula, four out of the six appearing terms obviously vanish and we are left with
\begin{align*}
\sff^F(\partial_{x^i}^{\sH M})(\partial_\alpha,\partial_\alpha) &
\stackrel{\hphantom{(21)}}{=} g_F\bigl([\partial_{x^i}^{\sH M},\partial_\alpha],\partial_\alpha\bigr) \\
&\stackrel{\eqref{eq:partialHM}}{=} g_F\Bigl(\underbrace{[\partial_i,\partial_\alpha]}_{=0} - \bigl[g_{i,d+\beta}(x,n)\partial_\beta,\partial_\alpha\bigr],\partial_\alpha\Bigr) \\
&\stackrel{\hphantom{(21)}}{=} \frac{g_{i,d+\beta}(x,n)}{\partial n^\alpha} \underbrace{g_F (\partial_\beta,\partial_\alpha)}_{\updelta_{\beta\alpha}} \\
&\stackrel{\hphantom{(21)}}{=} g_B^\bot(\omega^\sN(\partial_{x^i})e_\alpha,e_\alpha)\,.
\end{align*}
But now, the last expression equals zero since $\omega^\sN(\partial_{x^i})$ is $\fs\fo(k)$-valued. Consequently, the mean curvature vector $\eta_F$ defined by
\begin{equation}
\label{eq:etaF}
\tr_{\sT F} \sff(\partial_{x^i}^{\sH M}) = \pi^*_M g_B (\partial_{x^i}^{\sH M},\eta_F)
\end{equation}
vanishes identically. Finally note that the same considerations also hold for the submersion metric $g_\mathrm{s}^\eps$ due to $g^\eps|_{\sV M}=g_F = g^\eps_\mathrm{s}|_{\sV M}$. 
\end{Remark}

\subsection{The Horizontal Laplacian}\label{sect:Laplace}

Now that we have a detailed description of the metric,  we can explicitly express the horizontal Laplacian by the vector fields $(\{\partial_{x^i}^{\sH M}\}_{i=1}^d, \{\partial_{d+\alpha}\}_{\alpha=1}^k)$. Let $(\{\pi^*_M \dd x^i\}_{i=1}^d,\{\delta n^\alpha\}_{\alpha=1}^k)$ be the dual basis (note that in general $\delta n^\alpha \neq \dd n^\alpha$ since $\dd n^\alpha(\partial_{x^i}^{\sH M})\neq 0$). Then by definition
\begin{align*}
 \delta n^\alpha(\projHM \grad_{g_\mathrm{s}}\psi)&=0\, , \\
  \pi^*_M\dd x^i(\projHM \grad_{g_\mathrm{s}}\psi) &= g_\mathrm{s}^{jk}(\partial_{x^j}^{\sH M}\psi) \dd x^i(\partial_{x^k})=g_B^{ij}\partial_{x^j}^{\sH M}\psi 
\end{align*}
and thus
\[
\grad_{g_\mathrm{s}}\psi= g_B^{ij}(\partial_{x^i}^{\sH M}\psi) \partial_{x^j}^{\sH M}\, .
\]
When acting on a horizontal vector field $Y$, the divergence takes the coordinate form
\begin{align}
\dive_{g_\mathrm{s}}Y &= \frac{1}{\sqrt{\abs{g_\mathrm{s}}}}\partial_{x^i}^{\sH M}\sqrt{\abs{g_\mathrm{s}}}(\pi^*_M\dd x^i(Y)) \nonumber \\
&=\frac{1}{\sqrt{\abs{g_B}}}\partial_{x^i}^{\sH M}\sqrt{\abs{g_B}}\bigl(\pi^*_M\dd x^i(Y)\bigr) - \pi_M^* g_B(\eta_F, Y)\, , \label{eq:divgs}
\end{align}
since $\sqrt{\abs{g_F}}^{-1} \big(\partial_{x^i}^{\sH M} \sqrt{\abs{g_F}}\big)= 
-g_\mathrm{s}(\eta_F,\partial_{x^i}^{\sH M})$. For a horizontal lift $X^{\sH M}$ 
we have the simple formula
\[
 \dive_{g_\mathrm{s}} X^{\sH M}=\pi^*_M \bigl(\dive_{g_B}X - g_B(\pi_{M*}\eta_F, X)\bigr)\, .
\]
Now for a massive waveguide $\eta_F=0$ and the horizontal Laplacian takes the familiar form
\[
 \Delta_{\sH} = \frac{1}{\sqrt{\abs{g_B}}}\partial_{x^i}^{\sH M}\sqrt{\abs{g_B}}g_B^{ij} \partial_{x^j}^{\sH M}\, ,
\]
which is just $\Delta_{g_B}$ with $\partial_{x^i}$ replaced by $\partial_{x^i}^{\sH M}$.
\subsection{The Bending Potential}
In the introduction (see equation~\eqref{eq:Vb intro}) we saw that the leading order of $\Vb$ is attractive (negative) and proportional to the square of the curve's curvature $\kappa=\abs{c''}$. 
Here we give a detailed derivation of $\Vb$ for generalised massive waveguides and then discuss the sign of its leading part. Therefor, let $\{\tau_i\}_{i=1}^d$ be a local orthonormal frame of $\sT B$ with respect to $g_B$ and let $\{n^\alpha\}_{\alpha=1}^k$ be coordinates on $\sN B$ as in equation~\eqref{eq:xn}. Then
\[
T_i:=\tau_i^{\sH M}\, ,\quad N_\alpha:=\tfrac{\partial}{\partial n^\alpha}
\]
for $i\in\{1,\dots,d\}$ and $\alpha\in\{1,\dots,k\}$ form a local frame of $\sT M$. In this frame the scaled metrics have the form (see also~\eqref{eq:horblock})
\[
g^\eps = \begin{pmatrix}[c|c]
  \eps^{-2}(\id_{d\times d} -\eps \cW(\nu))^2 &  0\\ \hline
  0 & \id_{k\times k}
\end{pmatrix}
\,,\quad
g^\eps_\mathrm{s} = \begin{pmatrix}[c|c]
  \eps^{-2}\id_{d\times d} &  0\\ \hline
  0 & \id_{k\times k}
\end{pmatrix}.
\]
From that and equation~\eqref{eq:rhoepsgen} we easily conclude that
\[
\rho_\eps = \sqrt{\frac{\det(g^\eps)}{\det(g^\eps_\mathrm{s})}} = \det\bigl(\id_{d\times d} - \eps \cW(\nu)\bigr) = \exp\Bigl(\tr\log\bigl(\id_{d\times d} - \eps \cW(\nu)\bigr)\Bigr) .
\]
Using Taylor's expansion for $\eps$ small enough,
\[
-\log\bigl(\id_{d\times d} - \eps \cW(\nu)\bigr) = \underbrace{\eps \cW(\nu) + \tfrac{\eps^2}{2} \cW(\nu)^2 + \tfrac{\eps^3}{3} \cW(\nu)^3}_{=:\cZ(\eps)} + \cO(\eps^4)\, ,
\]
we have $
\log(\rho_\eps) = - \tr \cZ(\eps) + \cO(\eps^4)$.
Next, we calculate the terms appearing in $\Vb$~\eqref{eq:Vrhoeps} separately:
\[
\Delta_\sH \log \rho_\eps = - \eps \Delta_\sH\tr\cW(\nu) + \cO(\eps^2)\, ,
\]
\begin{align*}
\eps^{-2}\Delta_\sV \log \rho_\eps &= - \eps^{-2}\sum_{\alpha=1}^k \partial_{n^\alpha}^2 \tr \cZ(\eps) + \cO(\eps^2) \\
&= -\sum_{\alpha=1}^k \Bigl[\tr\bigl(\cW(e_\alpha)^2\bigr) + 2 \eps \tr\bigl(\cW(e_\alpha)^2\cW(\nu)\bigr)\Bigr] + \cO(\eps^2)\, ,
\end{align*}
\begin{align*}
\dd \log \rho_\eps &= \projHM \dd \log\rho_\eps - \tr\bigl(\partial_{n_\alpha} \cZ(\eps)\bigr)\, \dd n^\alpha + \cO(\eps^4) \\
&= \projHM \dd\log \rho_\eps - \tr\Bigl( \eps\cW(\e_\alpha)\bigl(\id_{d\times d}+\eps\cW(\nu) + \eps^2 \cW(\nu)^2\bigr)\Bigr) \, \dd n^\alpha + \cO(\eps^4)\, , 
\end{align*}
denoting by $\projHM$ the adjoint of the original $\projHM$ with respect to the pairing of $\sT^*M$ and $\sT M$, and hence
\begin{align*}
& \eps^{-2}g_F(\dd\log\rho_\eps,\dd\log\rho_\eps) \\
&= \sum_{\alpha=1}^k 
\Bigl[\bigl(\tr\cW(e_\alpha)\bigr)^2 + 2 \eps \tr\bigl(\cW(e_\alpha)\bigr) \tr\bigl(\cW(e_\alpha)\cW(\nu)\bigr)\Bigr] + \cO(\eps^2)\, .
\end{align*}
Putting all this together, we obtain the following expression for the bending potential in the case of massive quantum waveguides:
\begin{align}
\Vb &= \frac{1}{4} \sum_{\alpha=1}^k \Bigl[\bigl(\tr\cW(e_\alpha)\bigr)^2 - 2 \tr\bigl(\cW(e_\alpha)^2\bigr)\Bigr] \label{eq:Vbend_0}\\
&\hphantom{=}\ + \frac{\eps}{2} \sum_{\alpha=1}^k \Bigl[\tr\cW(e_\alpha) \tr\bigl(\cW(e_\alpha)\cW(\nu)\bigr) - 2\tr\bigl(\cW(e_\alpha)^2\cW(\nu)\bigr) - \Delta_\sH \tr \cW(\nu)\Bigr] \label{eq:Vrhoeps3} \\
&\hphantom{=}\ + \cO(\eps^2)\, . \nonumber 
\end{align}
The leading term of this expression ($\Vb^0:=\eqref{eq:Vbend_0}$) has been widely stressed in the literature concerning one-dimensional quantum waveguides (see e.g.~\cite{DE95, Kre07}), where it has a purely attractive effect. Its higher dimensional versions were discussed by Tolar~\cite{Tolar1988} but are generally less known, so we will discuss their possible effects for the rest of this section.

Since $\cW(e_\alpha)$ is self-adjoint, we may choose for each $\alpha\in\{1,\dots,k\}$ the orthonormal frame $\{\tau_i\}_{i=1}^d$ such that it consists of the eigenvectors of $\cW(e_\alpha)$ with eigenvalues (principal curvatures) $\{\kappa_i^\alpha\}_{i=1}^d$. In order to get an impression of $\Vb^0$'s sign, we divide $\cW(e_\alpha)$ into a traceless part $\cW_0(e_\alpha)$ and a multiple of the identity:
\[
\cW(e_\alpha) = \cW_0(e_\alpha) + \frac{H_\alpha}{d} \id_{d\times d}\, .
\]
Note that the prefactors $H_\alpha$ equal the components of the mean curvature vector of $B$ in direction $e_\alpha$. With the notation
\[
\norm{M}^2 := \tr\bigl(M^\mathrm{t} M\bigr)\geq 0
\]
for any $M\in\RR^{d\times d}$, we get the relation
\[
\norm{\cW(e_\alpha)}^2 = \norm{\cW_0(e_\alpha)}^2 + \frac{H_\alpha^2}{d}
\]
for all $\alpha\in\{1,\dots,k\}$ since $\cW_0(\cdot)$ is traceless. This yields for the potential~\eqref{eq:Vbend_0}:
\begin{align*}
\Vb^0 &= \frac{1}{4} \sum_{\alpha=1}^k\bigl[H_\alpha^2 - 2 \norm{\cW(e_\alpha)}^2\bigr] \\
&= \frac{1}{4} \sum_{\alpha=1}^k \left[H_\alpha^2 - 2 \left(\norm{\cW_0(e_\alpha)}^2 + \frac{H_\alpha^2}{d}\right)\right] \\
&= \frac{1}{4} \sum_{\alpha=1}^k \left[\left(1-\frac{2}{d}\right) H_\alpha^2 - 2\norm{\cW_0(e_\alpha)}^2\right].
\end{align*} 
The latter relation shows that for $d\in\{1,2\}$ the leading order of the bending potential is non-positive. Thus, the effect of bending has an attractive character ($\Vb^0<0$) for $\eps$ small enough, or is of lower order ($\Vb^0=0$), independently of the codimension $k$. For $d\geq 3$, the first term is non-negative and may overcompensate the second term leading to a positive contribution to $\Vb$. Consequently, a repulsive bending effect is possible.

\begin{Example}
We may rewrite expression~\eqref{eq:Vbend_0} in terms of principal curvatures as
\begin{equation}
\label{eq:Vbendkappa}
\Vb^0 = \frac{1}{4} \sum_{\alpha=1}^k \left[ \left(\sum_{i=1}^d \kappa_i^\alpha\right)^2 - 2 \sum_{i=1}^d (\kappa_i^\alpha)^2\right].
\end{equation}
\begin{enumerate}
\item For a waveguide modelled around a curve $c$, $d=1$, one immediately sees that $\Vb^0=-\tfrac14 \kappa^2=- \tfrac14 \abs{c''}^2$.
\item We consider the case where $B\subset \RR^{d+1}$ is the $d$-di\-men\-sion\-al standard sphere of radius $R$. The principal curvatures in the direction of the outer-pointing normal are given by $\kappa_i = 1/R$ for all $i\in\{1,\dots,d\}$, hence the bending potential~\eqref{eq:Vbendkappa} reads
\begin{align*}
\Vb^0 &= \frac{1}{4} \left[\left(\sum_{i=1}^d \frac{1}{R}\right)^2 - 2 \sum_{i=1}^d \left(\frac{1}{R}\right)^2\right] = \left(1-\frac{2}{d}\right)\frac{d^2}{4 R^2}\, .
\end{align*}
It follows that $\Vb^0 < 0$ for $d=1$, $\Vb^0 = 0$ for $d=2$ and $\Vb^0>0$ for $d\geq 3$, respectively. Thus, depending on the dimension $d$ of the sphere, the effect of bending can be either attractive or repulsive.
\end{enumerate}
\end{Example}

\subsection{The Adiabatic Hamiltonian}

We are now ready to calculate the geometric terms in the adiabatic operator. In this we concentrate on the adiabatic operator~\eqref{eq:Ha} and explicitly calculate all the relevant terms on the energy scale given by Theorem~\ref{thm:low spectrum}. First we take care of the contribution of $H_1$, then we turn to the potential $V_\mathrm{a}$ and explain its connection to \enquote{twisting} of the quantum waveguide.

\subsubsection{The Operator \texorpdfstring{$P_0H_1P_0$}{P\textzeroinferior H\textoneinferior P\textzeroinferior}}

The contribution of the bending potential, that was calculated in the previous section, is given by its adiabatic approximation
\[
\Vb^\mathrm{a}:=P_0 \Vb P_0= \int_{F_x}  \Vb(\nu) \abs{\phi_0(\nu)}^2\ \dd\nu \, .
\]
Since the leading part $\Vb^0$ is independent of the fibre coordinate $\nu$, it is unchanged by this projection. The next term in the expansion of $\Vb$ is given by~\eqref{eq:Vrhoeps3}.
The Weingarten map is linear in $\nu$ and since 
\[
\partial_{x^i}^{\sH M}n^\alpha \stackrel{\eqref{eq:partialHM}}{=}- g_B^\bot\bigl(\omega^\sN(\partial_{x^i})\nu, e_\beta\bigr) \partial_{n^\beta}n^\alpha = - g_B^\bot\bigl(\omega^\sN(\partial_{x^i})\nu, e_\alpha\bigr)
\]
is again linear in $\nu$, $\Delta_\sH\tr \mathcal{W}(\nu)$ is also linear in $\nu$. Consequently, the contribution of~\eqref{eq:Vrhoeps3} to $\Vb^\mathrm{a}$ is proportional to 
\[
\scpro{\phi_0}{\nu \phi_0}_{F_x} = \int_{F_x}  \nu \abs{\phi_0(\nu)}^2\ \dd\nu\, .
\]
Hence, this contribution vanishes if the centre of mass of the ground state $\phi_0$ lies exactly on the submanifold $B$. This is a reasonable assumption to make and represents a \enquote{correct} choice of parametrisation of the waveguide. Under this assumption we have
\[
\Vb^\mathrm{a}= \Vb^0 + \mathcal{O}(\eps^2)\, .
\]
From the expression~\eqref{eq:horblock} for the horizontal block of the metric  $g^\eps$ one obtains its expansion   on horizontal  one-forms by locally inverting the matrix $(g^\eps)_{ij}$ (see~\cite{Wit}). The result is 
\begin{equation}\label{eq:dual metric}
 g^\eps(\pi_M^*\dd x^i, \pi_M^*\dd x^j)=\eps^2 \big(g_B^{ij} + 2\eps \sff(\nu)^{ij} + \mathcal{O}(\eps^2)\big)\,,
\end{equation}
where $\sff$ denotes the second fundamental form of $B$, defined on $\sT^*B$ by $\sff^{ij}:=\sff_{kl}g_B^{ik}g_B^{jl}$.  Moreover,
we extend the latter to $\sT^*M$, understanding $\sff(\nu)$ as its lift to the horizontal part $\sH^*M$ and extending to $\sT^*M$ by zero.
The vertical components of $g^\eps$ and $g_\mathrm{s}^\eps$ coincide, hence as an operator on $L^2(B)$ we have the expression
\begin{equation}\label{eq:H1 massive nocent}
(P_0 S^\eps P_0)\psi=2 \int_{F_x} \phi_0 \dive_{g_\mathrm{s}} \Bigl(\sff(\nu)\bigl(\dd(\phi_0\psi), \cdot\bigr)\Bigr)\ \dd\nu + \mathcal{O}(\eps)
\end{equation}
with an error of order $\eps$ on $W^2(B)$. Using the Leibniz rule we can rewrite this as
\begin{align}
&2 \int_{F_x} 2 \phi_0 \sff(\nu)\big(\dd\phi_0, \dd \psi\big) + \abs{\phi_0}^2 \dive_{g_\mathrm{s}}\bigl(\sff(\nu)(\dd\psi, \cdot)\bigr) + \phi_0 \psi \dive_{g_\mathrm{s}}\bigl(\sff(\nu)(\dd\phi_0, \cdot)\bigr)\ \dd\nu\nonumber \\
 &\ =2\int_{F_x} \dive_{g_\mathrm{s}} \bigl(\abs{\phi_0}^2 \sff(\nu)(\dd\psi, \cdot)\bigr) +
 \phi_0 \psi \dive_{g_\mathrm{s}}\bigl(\sff(\nu)(\dd\phi_0, \cdot)\bigr)\ \dd\nu\, . \label{eq:H_1 allg}
\end{align}
Now $\phi_0$ vanishes on the boundary and 
$\sff(\dd\psi, \cdot)$ is a horizontal vector field, so by~\eqref{eq:divgs} we have
\begin{equation}
\label{eq:diff H_1 massive}
 \int_{F_x} \dive_{g_\mathrm{s}} \bigl(\abs{\phi_0}^2 \sff(\nu)(\dd\psi, \cdot)\bigr)\ \dd\nu
 =\dive_{g_B}\int_{F_x} \abs{\phi_0}^2 \sff(\nu)(\dd\psi, \cdot)\ \dd\nu \, .
\end{equation}
If we assume again that $\phi_0$ is centred on $B$, this term vanishes and we are left with the potential
\begin{equation}
\label{eq:H_1 massive}
\eps P_0 H_1 P_0= \eps^2 \Vb^0 - 2\eps^3 \int_{F_x} \phi_0 \dive_{g_\mathrm{s}}\bigl(\sff(\nu)(\dd\phi_0, \cdot)\bigr)\ \dd\nu +\mathcal{O}(\eps^4)
\end{equation}
with an error bound in $\mathcal{L}\big(W^2(B), L^2(B)\big)$.

\subsubsection{The Adiabatic Potential \texorpdfstring{$V_\mathrm{a}$}{} and \enquote{Twisted} Waveguides}\label{sect:twist}

Since the fibres $F_x$ are completely geodesic with respect to $g_F$ for massive quantum waveguides (\cf Remark~\ref{rem:complgeod}), we have $\eta_F=0$ and the adiabatic potential defined in~\eqref{eq:Va} reduces to 
\begin{equation}\label{eq:Va massive}
V_\mathrm{a} = \int_{F_x} \pi_M^* g_B(\grad_{g_\mathrm{s}}\phi_0, \grad_{g_\mathrm{s}}\phi_0)\ \dd\nu\, .
\end{equation}
This is called the Born-Huang potential in the context of the Born-Oppenheimer approximation.
This potential is always non-negative. It basically accounts for the alteration rate of $\phi_0$ in horizontal directions. 

In the literature, the adiabatic potential has been studied mainly for \enquote{twisted} quantum waveguides. 
These have two-dimensional fibres $F_x$ which are isometric but not invariant under rotations and twist as one moves along the one-dimensional base curve $B$~\cite{Kre07}. The operators $\Delta_{F_x}$, $x\in B$, are isospectral and their non-trivial dependence on $x$ is captured by $V_\mathrm{a}$.

We now generalise this concept to massive waveguides of arbitrary dimension and codimension and calculate the adiabatic potential for this class of examples.
In this context, a massive quantum waveguide $F\to M\xrightarrow{\pi_M} B$ is said to be only twisted at $x_0\in B$, if there exist a geodesic ball $U\subset B$ around  $x_0$ and a local orthonormal frame $\{f_\alpha\}_{\alpha=1}^k$ of $\sN B|_U$ such that
\[
\pi^{-1}_M(U) = \bigl\{n^\alpha f_\alpha(x):\ (n^1,\dots,n^k)\in F,~ x\in U\bigr\}\, .
\]
This exactly describes the situation that the cross-sections $(F_x, g_{F_x})$ are isometric to $F\subset\RR^k$, but may vary from fibre to fibre by an $\mathrm{SO}(k)$-transformation. Moreover, it follows that $\lambda_0$ is constant on $U$ and the associated eigenfunction $\phi_0$ is of the form $\phi_0(\nu(x)=n^\alpha f_\alpha(x))=\Phi_0(n^1,\dots,n^k)$, where $\Phi_0$ is the solution of
\[
-\Delta_n \Phi_0(n) = \lambda_0 \Phi_0(n)\, ,\quad \text{$\Phi_0(n)=0$ on $\partial F$}\, .
\]
As for the calculation of $V_\mathrm{a}$ at $x_0$, we firstly compute for $\partial_{x^i}^{\sH M}\in\Gamma(\sH M)$:
\begin{align}
\pi_M^* g_B(\grad_{g_\mathrm{s}} \phi_0, \partial_{x^i}^{\sH M})|_{\nu(x_0)} &\stackrel{\hphantom{\text{\eqref{eq:partialHM}}}}{=} g_\mathrm{s}(\grad_{g_\mathrm{s}} \phi_0, \partial_{x^i}^{\sH M})|_{\nu(x_0)} \nonumber \\
&\stackrel{\hphantom{\text{\eqref{eq:partialHM}}}}{=} \partial_{x^i}^{\sH M} \phi_0\bigr|_{\nu(x_0)} \nonumber \\
&\stackrel{\text{\eqref{eq:partialHM}}}{=} \Bigl[\partial_i - g_B^\bot\bigl(\omega^\sN(\partial_{x^i}) \nu, f_\beta\bigr)\bigr|_{x_0} \partial_{n^\beta}\Bigr] \Phi_0(n) \nonumber \\
&\stackrel{\hphantom{\text{\eqref{eq:partialHM}}}}{=} - n^\alpha g_B^\bot \bigl(\nabla^\sN_{\partial_{x^i}} f_\alpha, f_\beta\bigr)\bigr|_{x_0} \frac{\partial\Phi_0(n)}{\partial n^\beta}\, . \label{eq:twist1}
\end{align}
In order to get a better understanding of $g_B^\bot(\nabla^\sN_{\partial_{x^i}} f_\alpha, f_\beta)|_{x_0}$, we introduce 
on $U$ a locally untwisted  orthonormal frame $\{e_\alpha\}_{\alpha=1}^k$ of $\sN B|_U$.
It is obtained by taking the vectors $f_\alpha(x_0)\in \sN_{x_0} B$ and parallel transporting them along radial geodesics with respect to the normal connection $\nabla^{\sN}$. Thus, twisting is always to be understood relative to the locally parallel frame $\{e_\alpha\}_{\alpha=1}^k$.
The induced   map  that transfers the reference frame $\{e_\alpha\}_{\alpha=1}^k$ into the twisting frame $\{f_\alpha\}_{\alpha=1}^k$ is denoted by $R:U\to\mathrm{SO}(k)$. It is defined by the relation $f_\alpha(x) = e_\gamma(x) R^\gamma_{\hphantom{\gamma}\alpha}(x)$ for $x\in U$ and obeys $R(x_0)=\id_{k\times k}$ due to the initial data of $\{e_\alpha\}_{\alpha=1}^k$. Consequently, using the differential equation of the parallel transport, we have
\[
\nabla^\sN_{\partial_{x^i}} f_\alpha(x_0) = \nabla^\sN_{\partial_{x^i}}(e_\gamma R^\gamma_{\hphantom{\gamma}\alpha})(x_0)= \underbrace{\big(\nabla^\sN_{\partial_{x^i}} e_\gamma\big) (x_0)}_{=0} \updelta^\gamma_{\hphantom{\gamma}\alpha} + e_\gamma(x)\, \partial_{x^i} R^\gamma_{\hphantom{\gamma}\alpha}(x_0)
\]
and hence
\begin{equation}
\label{eq:twist2}
g_B^\bot \bigl(\nabla^\sN_{\partial_{x^i}} f_\alpha, f_\beta\bigr)\bigr|_{x_0} = g_B^\bot(e_\gamma \,\partial_{x^i} R^\gamma_{\hphantom{\gamma}\alpha}, e_\beta)|_{x_0} = \partial_{x^i} R_{\beta\alpha}(x_0)\, .
\end{equation}
For $1\leq\alpha<\beta\leq k$, let $T_{\alpha\beta}\in\RR^{k\times k}$ defined by
\[
(T_{\alpha\beta})_{\gamma\zeta} := \updelta_{\alpha\zeta}\updelta_{\beta\gamma} - \updelta_{\alpha\gamma}\updelta_{\beta\zeta}
\]
be a set of generators of the Lie Algebra $\fs\fo(k)$. This induces generalised angle functions $\{\omega^{\alpha\beta}\in C^\infty(U)\}_{\alpha<\beta}$ by the relation
\[
R(x) = \exp\Big(\sum_{\alpha<\beta} \omega^{\alpha\beta}(x) T_{\alpha\beta}\Big)
\]
for $x\in U$. Then a short calculation shows that
\begin{equation}
\label{eq:twist3}
\partial_{x^i} R(x_0) = \bigl((\partial_{x^i}\omega^{\alpha\beta}T_{\alpha\beta}) R \bigr)(x_0) = \dd\omega^{\alpha\beta}(\partial_{x^i})|_{x_0} T_{\alpha\beta}
\end{equation}
for $\alpha<\beta$. Combining~\eqref{eq:twist1},~\eqref{eq:twist2} and~\eqref{eq:twist3}, we obtain
\[
\pi_M^* g_B(\grad_{g_\mathrm{s}} \phi_0, \partial_{x^i}^{\sH M})|_{\nu(x_0)=n^\alpha f_\alpha(x_0)} = - \dd\omega^{\alpha\beta}(\partial_{x^i})|_{x_0} (L_{\alpha\beta} \Phi_0)(n)\, ,\quad \alpha<\beta\, ,
\]
where
\[
L_{\alpha\beta}: \Phi_0(n)\mapsto (L_{\alpha\beta}\Phi_0)(n) := \scpro{(\nabla_n \Phi_0)(n)}{T_{\alpha\beta} n}_{\RR^k} = (n^\alpha \partial_{n^\beta} - n^\beta \partial_{n^\alpha}) \Phi_0(n)
\]
defines the action of the $(\alpha,\beta)$-component of the angular momentum operator in $k$ dimensions. From here, it is easy to see that the adiabatic potential at $x_0$ is given by
\begin{align}
V_\mathrm{a}(x_0) &= \int_{F_{x_0}} \pi_M^* g_B(\grad_{g_\mathrm{s}} \phi_0, \grad_{g_\mathrm{s}} \phi_0)\ \dd \nu \nonumber \\
&= \int_F g_B \bigl( - \dd\omega^{\alpha\beta} (L_{\alpha\beta}\Phi_0)(n),- \dd\omega^{\gamma\zeta} (L_{\gamma\zeta}\Phi_0)(n)\bigr)\bigr|_{x_0}\ \dd n \nonumber \\
&= \underbrace{g_B (\dd\omega^{\alpha\beta},\dd\omega^{\gamma\zeta})|_{x_0}}_{=:\mathbf{  R}^{(\alpha\beta),(\gamma\zeta)}(x_0)} \underbrace{\scpro{L_{\alpha\beta}\Phi_0}{L_{\gamma\zeta} \Phi_0}_{L^2(F)}}_{=: \mathbf{L }_{(\alpha\beta),(\gamma\zeta)}}\, ,\quad \text{$\alpha<\beta$ and $\gamma<\zeta$} \label{eq:Vtwist} \\
&= \tr_{\RR^{k(k-1)/2}} \bigl(\mathbf{  R}(x_0)^\mathrm{t} \mathbf{L }\bigr)\, . \nonumber
\end{align}
The first matrix $\mathbf{ R}(x_0)$ encodes the rate, at which the frame $\{f_\alpha\}_{\alpha=1}^k$ twists relatively to the parallel   frame $\{e_\alpha\}_{\alpha=1}^k$ at $x_0$.   The second matrix $\mathbf{L }$ measures the deviation of  the eigenfunction $\Phi_0$  from being rotationally invariant.
It determines to which extent   the twisting of the waveguide effects the states in the range of $P_0$ 
 and it depends only on the set $F\subset\RR^k$ (and not on the point $x_0$ of the submanifold $B$). Finally for the case of a twisted quantum waveguides with $(B,g_B)\cong (\RR,\updelta^1)$ and $k=2$, there exists only one angle function $\omega\in C^\infty(\RR)$ and one angular momentum operator $L=n^1 \partial_{n^2} - n^2 \partial_{n^1}$. Then formula~\eqref{eq:Vtwist} yields the well-known result~\cite{KS12}
\[
V_\mathrm{a} = (\omega')^2 \norm{L\Phi_0}^2_{L^2(F)}\,,
\]
which clearly vanishes if $F$ is invariant under rotations.
\subsubsection{Conclusion}\label{sect:H_a alpha}
Now that we have calculated all the relevant quantities, we can give an explicit expansion of $H_\mathrm{a}$.
The correct norm for error bounds of course depends on the energy scale under consideration. For a constant eigenvalue $\lambda_0$ and $\alpha=2$ the graph-norm of $\eps^{-2} H_\mathrm{a}$ is clearly equivalent (with constants independent of $\eps$) to the usual norm of $W^2(B,g_B)$. In this situation the best approximation by $H_\mathrm{a}$ given by Theorem~\ref{thm:low spectrum} has errors of order $\eps^4$, so the estimates just derived give
\begin{equation}
\label{eq:X3}
\tag{$\alpha=2$}
 H_\mathrm{a} = -\eps^2\Delta_{g_B} + \lambda_0 + \eps^2 V_\mathrm{a} + \eps^2\Vb^0 -  2 \eps^3  \int_{F_x} \phi_0 \dive_{g_\mathrm{s}}\bigl(\sff(\nu)(\dd\phi_0, \cdot)\bigr)\ \dd\nu + \mathcal{O}(\eps^4)
\end{equation}
if $\phi_0$ is centred. If this is not the case, the expansion can be read off from equations~\eqref{eq:Vrhoeps3},~\eqref{eq:H1 massive nocent} and~\eqref{eq:Va massive}.

If $\lambda_0$ has a non-degenerate minimum and $\alpha=1$, the errors of our best approximation are of order $\eps^3$. Thus, the potentials of order $\eps^3$ can be disregarded in this case. Note however that on the domain of $\eps^{-1}H_{\mathrm{a}}$ we have $\eps\partial_{x^i}^{\sH M}=\mathcal{O}(\sqrt\eps)$, so the differential operator~\eqref{eq:diff H_1 massive} will be relevant. The error terms of equation~\eqref{eq:H_1 massive}, containing second order differential operators, are of order $\eps^3$ with respect to $\eps^{-1}H_\mathrm{a}$, so they are still negligible. Thus, for $\psi \in W^2(B)$ with $\norm{\psi}^2 + \norm{(-\eps\Delta_{g_B} + \eps^{-1}\lambda_0)\psi}^2=\mathcal{O}(1)$ we have
\begin{equation}
\label{eq:X4}
\tag{$\alpha=1$}
H_\mathrm{a}\psi= \bigl(-\eps^2\Delta_{g_B} + \lambda_0 + \eps^2 V_\mathrm{a} + \eps^2\Vb^0\bigr)\psi
 - 2\eps^3 \dive_{g_B}\int_{F_x} \abs{\phi_0(\nu)}^2 \sff(\nu)(\dd\psi, \cdot)\ \dd\nu
 + \mathcal{O}(\eps^3)\, ,
\end{equation}
where the last term is of order $\eps^2$ in general and vanishes for centred $\phi_0$.
\section{Hollow Quantum Waveguides} \label{chap:hollowQWG}
In this section we consider hollow quantum waveguides $F\to(M,g^\eps)\xrightarrow{\pi_M} (B,g_B)$, which by Definition~\ref{def:MassiveHollow} are the boundaries of massive waveguides. This underlying  massive wavguide  is denoted by $\mathring F\to (\mathring{M},\mathring{g}^\eps)\xrightarrow{\pi_{\mathring{M}}} (B,g_B)$ in the following. The bundle structure is inherited from the massive waveguide as well, \ie  $F = \partial \mathring{F}$  and the diagram
\begin{diagram}
 M   &\rInto    & \mathring{M}   \\
 \dTo^{\pi_M}    &&    \dTo_{\pi_{\mathring{M}}}\\
 B    &\rTo_{\id_B}&B
\end{diagram}
commutes.

Hollow quantum waveguides have, to our knowledge, not been studied before. In fact, already the derivations of $g^\eps$ and $\Vb$ constitute novel results. A slight generalisation of these calculations to objects that are not necessarily boundaries can be found in~\cite[Chapter 3]{Lampart2013}.

In order to determine the adiabatic operator $H_\mathrm{a}$ for hollow quantum waveguides, we follow the same procedure as layed out in the introduction and in the previous section. 

\subsection{The Pullback Metric}

Note that  $g^\eps=\mathring{g}^\eps|_{\sT M}$ and that we computed the unscaled pullback metric $\mathring{g}^{\eps=1}$ for the massive waveguide $\mathring{M}$ already in Lemma~\ref{lem:pullback}. The latter reads
\[
\mathring{g} := \mathring{g}^{\eps=1} = \underbrace{\pi_{\mathring{M}}^* g_B + \mathring{h}^{\eps=1}}_{=:{\mathring{g}}^\mathrm{hor}} + g_{\mathring{F}}\,,
\]
where the \enquote{horizontal correction} $\mathring{h}^{\eps=1}$~\eqref{eq:heps} vanishes on vertical vector fields and essentially depends on the extrinsic geometry of the embedding $B\hookrightarrow \RR^{d+k}$.

If we restrict $\mathring{M}$'s tangent bundle to $M$, one has the orthogonal decomposition
\begin{equation}
\label{eq:HVN}
\sT \mathring{M}|_M = \sT M \oplus \sN M \stackrel{\eqref{eq:THVM}}= \sH M \oplus \sV M \oplus \sN M 
\end{equation}
with respect to $\mathring{g}$.
Due to the commutativity of the above diagram, it follows that $\sV M \subset \sV \mathring{M}|_M$. This suggests to introduce the notation $\sV M^\bot$ for the orthogonal complement of $\sV M$ in $\sV \mathring{M}|_M$ with respect to $g_{\mathring{F}}$, \ie
\begin{equation}
\label{eq:decompgF}
\sV \mathring{M}|_M = \sV M \oplus \sV M^\bot\, .
\end{equation}
For any $X\in\Gamma(\sT B)$, let $X^{\sH {\mathring{M}}}\in\Gamma(\sH \mathring{M})$ and $X^{\sH M}\in\Gamma(\sH M)$ be the respective unique horizontal lifts. It then holds that
\[
\pi_{\mathring{M}*} \bigl(X^{\sH M} - X^{\sH {\mathring{M}}}|_M\bigr) = \pi_{M*} X^{\sH M} - \pi_{\mathring{M}*} X^{\sH {\mathring{M}}}|_M = X - X = 0\, .
\]
Thus, the difference between $X^{\sH M}$ and $X^{\sH {\mathring{M}}}|_M$ is a vertical field:
\begin{equation}
\label{eq:VX}
X^{\sH M} = X^{\sH {\mathring{M}}}|_M + V_X
\end{equation}
with $V_X\in \Gamma(\sV \mathring{M}|_M)$. Moreover, $V_X\in \Gamma(\sV M^\bot)$ since for arbitrary $W\in \Gamma(\sV M)\subset \Gamma(\sV \mathring{M}|_M)$
\[
0 = g(X^{\sH M},W) = \underbrace{\mathring{g}\bigl(X^{\sH {\mathring{M}}}|_M, W\bigr)}_{=0} + g_{\mathring{F}}\bigl(V_X,W)
\]
implies $g_{\mathring{F}}(V_X,W)=0$.
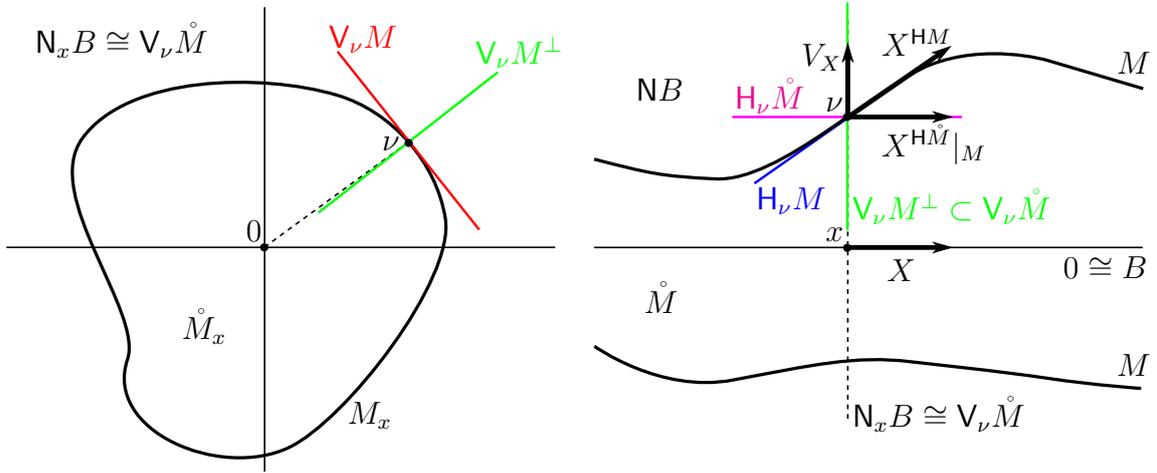
\begin{figure}[H]
\begin{center}
\includegraphics[width=0.95\textwidth]{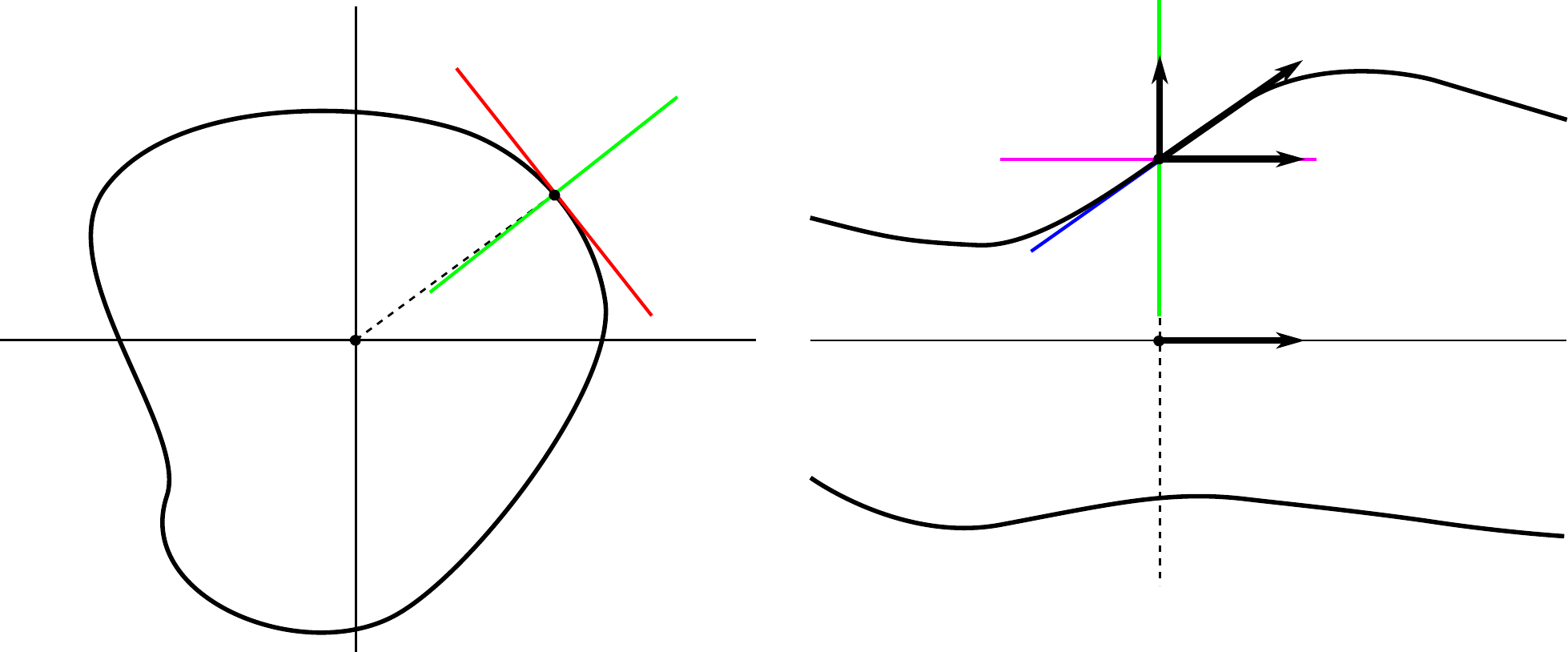}
\put(-10,36){$M$}
\put(-10,150){$M$}
\put(-96,72){$X$}
\put(-98,157){$X^{\sH M}$}
\put(-98,119){$X^{\sH {\mathring{M}}}|_M$}
\put(-128,153){$V_X$}
\put(-119,136){$\nu$}
\put(-119,87){$x$}
\put(-109,17){$\sN_xB \cong \sV_\nu \mathring{M}$}
\put(-190,140){$\sN B$}
\put(-187,60){$\mathring{M}$}
\put(-30,73){$0\cong B$}
\put(-153,137){\textcolor{magenta}{$\sH_\nu \mathring{M}$}}
\put(-145,99){\textcolor{blue}{$\sH_\nu M$}}
\put(-109,95){\textcolor{green}{$\sV_\nu M^\bot\subset \sV_\nu \mathring{M}$}}
\put(-250,154){\textcolor{green}{$\sV_\nu M^\bot$}}
\put(-305,160){\textcolor{red}{$\sV_\nu M$}}
\put(-285,121){$\nu$}
\put(-336,87){$0$}
\put(-298,18){$M_x$}
\put(-360,50){$\mathring{M}_x$}
\put(-415,158){$\sN_x B \cong \sV_\nu \mathring{M}$}
\end{center}
\caption{Left: Sketch of the fibre $\sN_x B$ for any $x\in B$. Note that for any $\nu\in \mathring{M}_x\subset\sN_xB$ we have the canonical identification of $\sN_xB$ and $\sV_\nu\mathring{M}$ via the isomorphism~\eqref{eq:isoalpha}. Right: Relationship between the horizontal lifts $X^{\sH M}$ and $X^{\sH \mathring{M}}|_M$. They are connected by the vertical field $V_X$.}
\label{fig:hollowQWG}
\end{figure}
\noindent Obviously, the relation $\pi_{M*} X^{\sH M}= X$ does not shed light on the vertical part $V_X$. The latter will be determined by the requirement $X^{\sH M}$ to be a tangent vector field on $M$, or equivalently by the condition $\mathring{g}(X^{\sH M},n)=0$, where $n\in\Gamma(\sN M)$ denotes a unit normal field of $M$ in $\mathring{M}$. In order to determine $V_X$ from this condition, we first need to show that the vertical component of $n$ is non-zero everywhere.
\begin{Lemma} \label{lem:v}
Let $n\in\Gamma(\sN M)$ be a unit normal field of the hollow quantum waveguide $M$. Then $v_n:=\projVrM n\in \Gamma(\sV M^\bot)$ is a non-vanishing vector field.
\end{Lemma}
\begin{Proof}
Decompose $n = v_n + h_n$ with $h_n:=\projHrM n\in\Gamma(\sH \mathring{M}|_M)$. It then holds for any vector field $W\in\Gamma(\sV M)$:
\[
g_{\mathring{F}}(W,v_n) = \mathring{g}(W,v_n) = \mathring{g}(W,v_n + h_n) = \mathring{g}(W,n) = 0\,,
\]
where we used~\eqref{eq:HVN} for the second and fourth equality. This clearly implies $v_n\in\Gamma(\sV M^\bot)$ by~\eqref{eq:decompgF}.
Now suppose there exists $\nu\in M$ with $v_n(\nu)=0$. Consider the space
\[
U_\nu := \sH_\nu M \oplus \spann\{(n(\nu)\} \subset \sT_\nu \mathring{M}\, .
\]
Since $n(\nu)\in \sN_\nu M$ is orthogonal to $\sH_\nu M\subset \sT_\nu M$, one has $\dim(U_\nu) = d+1$. We will show that the kernel of $\pi_{\mathring{M}*}|_{U_\nu}:U_\nu\to \im(\pi_{\mathring{M}*}|_{U_\nu})\subset \sT_{\pi_{\mathring{M}}(\nu)}B$ is trivial. Hence,
\[
d+1 = \dim(U_\nu) = \rank(\pi_{\mathring{M}*}|_{U_\nu})\leq \dim\bigl(\sT_{\pi_{\mathring{M}}(\nu)}B\bigr)
\]
clearly contradicts the fact that $\dim(B)=d$ and finally the assumption that $n(\nu)=0$. Therefor, let $w\in\ker(\pi_{\mathring{M}*}|_{U_\nu})\in \sV_\nu \mathring{M}\cap U_\nu$. On the one hand, since
\[
n(\nu)=\underbrace{v_n(\nu)}_{=0} + h_n(\nu)\in \sH_\nu \mathring{M} = \ker(\pi_{\mathring{M}*}|_\nu\bigr)^\bot\, ,
\]
$w$ is an element of $\sH_\nu M$. But on the other hand, $\pi_{\mathring{M}*}|_{\sH_\nu M}:\sH_\nu M\to \sT_{\pi_{\mathring{M}}(\nu)} B$ posseses a trivial kernel. Together, this yields $w=0$, \ie $\ker(\pi_{\mathring{M}*}|_{U_\nu})=\{0\}$.
\end{Proof}
In view of equation~\eqref{eq:VX}, Lemma~\ref{lem:v} suggests to define a function $\gimel(X)\in C^\infty(M)$ such that $V_X = \gimel(X)v_n$. Thus, the requirement $X^{\sH M}\in\Gamma(\sT M)$ yields
\begin{align*}
0 &= \mathring{g}(X^{\sH M},n) = \mathring{g}\bigl(X^{\sH {\mathring{M}}}|_M,n\bigr) + \mathring{g}(v_n,n) \gimel(X) \\
&= \mathring{g}^\mathrm{hor}\bigl(X^{\sH {\mathring{M}}}|_M,h_n\bigr) + g_{\mathring{F}}(v_n,v_n) \gimel(X)\, ,
\end{align*}
consequently
\begin{equation}
\label{eq:gimel}
\gimel(X) = -\frac{\mathring{g}^\mathrm{hor}({X^{\sH {\mathring{M}}}|}_M,h_n)}{g_{\mathring{F}}(v_n,v_n)}\, .
\end{equation}
Note that $\gimel(X)$ is well-defined since $g_{\mathring{F}}(v_n,v_n)>0$ by Lemma~\ref{lem:v}. Moreover, the latter equation shows that $\gimel\in\cT^0_1(B)\otimes C^\infty(M)$ is actually a tensor.

In summary, we just showed that the unscaled pullback metric on $M$ may be written as
\[
g= g^\mathrm{hor} + g_F\, ,\quad g_F:={g_{\mathring{F}}|}_{\sV M}
\]
with \enquote{horizontal block}
\[
g^\mathrm{hor}(X^{\sH M},Y^{\sH M}) := \mathring{g}^\mathrm{hor}\bigl(X^{\sH {\mathring{M}}}|_M,Y^{\sH {\mathring{M}}}|_M\bigr) +  g_{\mathring{F}}(v_n,v_n) \gimel(X)\gimel(Y)
\]
for $X,Y\in\Gamma(\sT B)$.
Going over to the scaled pullback metric $g^\eps$, we first show that the horizontal lift remains unchanged.
\begin{Lemma} \label{lem:hor}
Let $(M,g^\eps)\to(B,g_B)$ be a hollow quantum waveguide for $\eps>0$. Then the horizontal subbundle $\sH M$ is independent of $\eps$.
\end{Lemma}
\begin{Proof}
It is sufficient to show that for any vector field $X\in\Gamma(\sT B)$ its unique horizontal lift $X^{\sH M}$ is given by the $\eps$-independent expression
\[
X^{\sH M} = X^{\sH {\mathring{M}}}|_M + \gimel(X) v_n
\]
with $\gimel(X)\in C^\infty(M)$ and $v_n\in\Gamma(\sV M^\bot)$ as before. We already know that $X^{\sH M}$ is tangent to $M$ and satisfies $\pi_{M*} X^{\sH M}=X$. Thus, the requirement that $X^{\sH M}$ is orthogonal to any $W\in\Gamma(\sV M)$ with respect $g^\eps$ is the only possible way for any $\eps$-dependence to come into play. Therefore, we calculate
\begin{align*}
g^\eps(X^{\sH M},W) &= \mathring{g}^\eps \bigl( X^{\sH {\mathring{M}}}|_M + \gimel(X) v_n, W\bigr) \\
&= \underbrace{\eps^{-2}\Bigl[\pi_{\mathring{M}}^* g_B \bigl(X^{\sH {\mathring{M}}}|_M, W\bigr) + \eps \mathring{h}^\eps \bigl(X^{\sH {\mathring{M}}}|_M,W\bigr)\Bigr]}_{\text{$=0$, since $W\in\Gamma(\sV M)\subset\Gamma(\sV \mathring{M}|_M)$}} + \gimel(X) \underbrace{g_{\mathring{F}}(v_n,W)}_{\text{$=0$ by~\eqref{eq:decompgF}}} \\
&=0\, .
\end{align*}
\end{Proof}
In summary, if the scaled pullback metric of the massive waveguide $\mathring{M}$ has the form $\mathring{g}^\eps = \eps^{-2}(\pi^*_{\mathring{M}} g_B + \eps \mathring{h}^\eps) + g_{\mathring{F}}$, the scaled pullback metric $g^\eps$ of the associated hollow waveguide $M$ reads
\begin{equation}\label{eq:g^eps hollow}
g^\eps = \eps^{-2}(\pi^*_M g_B + \eps h^\eps) + g_F
\end{equation}
with
\[
h^\eps(X^{\sH M},Y^{\sH M}) := \mathring{h}^\eps\bigl(X^{\sH {\mathring{M}}}|_M,Y^{\sH {\mathring{M}}}|_M\bigr) + \eps g_{\mathring{F}}(v_n,v_n) \gimel(X)\gimel(Y)
\]
for $X,Y\in\Gamma(\sT B)$. This shows that the scaled pullback metric $g^\eps$ is again of the form~\eqref{eq:formg}.
\begin{Example}
Let us consider a simple example of a hollow quantum waveguide with $d=1$, $k=2$. Take $B=\{(x,0,0)\in\RR^3:\ x\in\RR\}\subset \RR^3$ as submanifold and parametrise the according massive quantum waveguide via
\[
\mathring{M} := \Bigl\{(x,0,0) + \varrho\, r(x,\varphi) e_r:\ (x,\varphi,\varrho)\in\RR\times[0,2\uppi)\times[0,1]\Bigr\} ,
\]
where $r:\RR\times[0,2\uppi)\to [r_-,r_+]$ with $0<r_-<r_+<\infty$ is a smooth function obeying the periodicity condition $r(\cdot,\varphi+2\uppi) = r(\cdot,\varphi)$ and $e_r=(0,\cos\varphi,\sin\varphi)\in\sV_{(x,\varphi,\varrho)} \mathring{M}$ stands for the \enquote{radial unit vector}. In view of Example~\ref{ex:conv} with $\kappa\equiv 0$, the unscaled pullback metric on $\mathring{M}$ is given by
\[
\mathring{g} = \mathring{g}^\mathrm{hor} + g_{\mathring{F}} = \dd x^2 + (\varrho^2\, \dd\varphi^2 + \dd\varrho^2)\, .
\]
Furthermore, we immediately observe that $\sT_x B=\spann\{\partial_x\}$ with trivial horizontal lift $\partial_x^{\sH {\mathring{M}}}=(1,0,0)=:e_x\in\sH_{(x,\varphi,\varrho)}\mathring{M}$. The hollow quantum waveguide associated to $M$ is obviously given by
\[
M := \Bigl\{(x,0,0) + r(x,\varphi) e_r:\ (x,\varphi)\in\RR\times[0,2\uppi)\Bigr\} = \mathring{M}|_{\varrho=1}\, .
\]
Consequently, $\sT_{(x,\varphi)}M$ is given by $\spann\{\tau_x,\tau_\varphi\}$, where
\begin{align*}
\tau_x(x,\varphi) &= \tfrac{\partial M}{\partial x}(x,\varphi) = e_x + \tfrac{\partial r}{\partial x} e_r\, , \\
\tau_\varphi(x,\varphi) &= \tfrac{\partial M}{\partial\varphi}(x,\varphi) = \tfrac{\partial r}{\partial \varphi} e_r + r e_\varphi
\end{align*}
with $e_\varphi=(0,-\sin\varphi,\cos\varphi)\in\sV_{(x,\varphi,\varrho)}\mathring{M}$. One easily agrees that $\tau_x$ and $\tau_\varphi$ are orthogonal to
\[
\tilde{n} = - \tfrac{\partial r}{\partial\varphi} e_\varphi + r e_r - r \tfrac{\partial r}{\partial x} e_x
\]
with respect to $\mathring{g}$. Hence,
\[
n(x,\varphi) := \frac{\tilde{n}}{\norm{\tilde{n}}_{\mathring{g}}} = \underbrace{\frac{- \frac{\partial r}{\partial\varphi} e_\varphi + r e_r}{\sqrt{(\frac{\partial r}{\partial\varphi})^2 + r^2\left[1+(\frac{\partial r}{\partial x})^2\right]}}}_{=:v_n\in\sV_{(x,\varphi)}M^\bot}  + \underbrace{\frac{- r\frac{\partial r}{\partial x} e_x}{\sqrt{(\frac{\partial r}{\partial\varphi})^2 + r^2\left[1+(\frac{\partial r}{\partial x})^2\right]}}}_{=:h_n\in\sH_{(x,\varphi)}\mathring{M}}
\]
is a unit normal vector of $M$ at $(x,\varphi)$ for $\eps=1$.
Noting that
\[
g_{\mathring{F}}(v_n,v_n) = \frac{(\frac{\partial r}{\partial\varphi})^2+ r^2}{(\frac{\partial r}{\partial\varphi})^2 + r^2\left[1+(\frac{\partial r}{\partial x})^2\right]}\, ,
\]
equation~\eqref{eq:gimel} gives
\begin{align*}
\gimel(\partial_x) &= -\frac{\mathring{g}^\mathrm{hor}(\partial_x^{\sH {\mathring{M}}}|_M,h_n)}{g_{\mathring{F}}(v_n,v_n)} \\
&= - \frac{- r\frac{\partial r}{\partial \varphi}}{\sqrt{(\frac{\partial r}{\partial\varphi})^2 + r^2\left[1+(\frac{\partial r}{\partial x})^2\right]}} \left(\frac{(\frac{\partial r}{\partial\varphi})^2+ r^2}{(\frac{\partial r}{\partial\varphi})^2 + r^2\left[1+(\frac{\partial r}{\partial x})^2\right]}\right)^{-1}  \\
&= \frac{r \frac{\partial r}{\partial\varphi}\sqrt{(\frac{\partial r}{\partial\varphi})^2 + r^2\left[1+(\frac{\partial r}{\partial x})^2\right]}}{(\frac{\partial r}{\partial \varphi})^2 + r^2}\, .
\end{align*}
This yields the following expression for the \enquote{horizontal block} of the scaled pullback metric $g^\eps$:
\begin{align*}
g^{\eps,\mathrm{hor}}(\partial_x^{\sH M},\partial_x^{\sH M}) &= \eps^{-2}\, \dd x^2\bigl(\partial_x^{\sH {\mathring{M}}}|_M,\partial_x^{\sH {\mathring{M}}}|_M\bigr) + g_{\mathring{F}}(v_n,v_n) \gimel(\partial_x)\gimel(\partial_x)  \\
&= \eps^{-2} + \frac{r^2(\frac{\partial r}{\partial x})^2}{(\frac{\partial r}{\partial\varphi})^2 + r^2} \\
&= \eps^{-2}\bigl(1 + \eps h^\eps(\partial_x^{\sH M},\partial_x^{\sH M}) \bigr)
\end{align*}
with
\[
h^\eps(\partial_x^{\sH M},\partial_x^{\sH M}) = \eps \frac{r^2(\frac{\partial r}{\partial x})^2}{(\frac{\partial r}{\partial\varphi})^2 + r^2}\, .
\]
\end{Example}
\subsection{The Adiabatic Hamiltonian}\label{sect:H_a hollow}
We now calculate the adiabatic operator for hollow waveguides. Since in this case the fibre is a manifold without boundary, the ground state of $H_F$ is explicitly known:
\[
\phi_0=\sqrt{\frac{\rho_\eps}{\norm{\rho_\eps}_1}}=\pi_M^*\mathrm{Vol}(F_x)^{-1/2} + \mathcal{O}(\eps)\, ,
\]
where $\norm{\rho_\eps}_1(x)$ is the $L^1$-norm of $\rho_\eps$ on the fibre $F_x$. Because of this we can express many of the terms appearing in $H_\mathrm{a}$, given in equation~\eqref{eq:Ha}, through $\rho_\eps$.

Let us begin with the sum of the modified bending potential $\tilde{V}_\mathrm{bend}$ appearing in $H_1$ and the adiabatic potential $V_\mathrm{a}$. First we obtain an expression for the one-form $\bar\eta$ by observing that for any vector field $X$ on $B$ 
\[
 0=X \int_{F_x} \abs{\phi_0}^2 \ \dd g_F = \int_{F_x} X^{\sH M} \abs{\phi_0}^2 \ \dd g_F - \underbrace{\int_{F_x} \abs{\phi_0}^2 g_B(X,\pi_{M*}\eta_F)\ \dd g_F}_{=\bar\eta(X)}\, . 
\]
So we see that
\begin{equation}
\label{eq:bareta}
 \bar\eta = \int_{F_x} \projHM \bigl(\dd\sabs{\phi_0}^2\bigr)\ \dd g_F\, .
\end{equation}
Now to start with the first term of the adiabatic potential can be calculated as in~\eqref{eq:diff H_1 massive}
\begin{align*}
\tr_{g_B} (\nabla^B \bar\eta)&\stackrel{\hphantom{(25)}}{=}\dive_{g_B} g_B(\bar \eta, \cdot)\\
&\stackrel{\eqref{eq:divgs}}{=}  \int_{F_x} \dive_{g_\mathrm{s}}\bigl(\projHM\grad_{g_\mathrm{s}}\abs{\phi_0}^2\bigr)\ \dd g_F\\
&\stackrel{\,\,\eqref{eq:LaplaceH}\,\,}{=}\int_{F_x}\Delta_\sH \abs{\phi_0}^2 \ \dd g_F\, .
\end{align*}
For the the modified bending potential one has, using the shorthand $\abs{F_x}=\mathrm{Vol}(F_x)$,
\begin{align*}
\eps^{2}\tilde{V}_\mathrm{bend}^\mathrm{a} &:=P_0 \big(\Vb - \tfrac{1}{2} \Delta_\sV (\log \rho_\eps) - \tfrac{1}{4} g_F(\dd\log\rho_\eps,\dd\log\rho_\eps)\big)P_0\\
  &\hphantom{:}=\frac{\eps^2}{2} \int_{F_x} \abs{\phi_0}^2 \Delta_\sH (\log\rho_\eps) \ \dd g_F+ \mathcal{O}(\eps^4)\\
  &\hphantom{:}=\frac{\eps^2}{2} \int_{F_x} \vert F_x \vert^{-1} (\Delta_\sH \rho_\eps) \ \dd g_F + \mathcal{O}(\eps^4)\, .
\end{align*}
Note that this expression is of order $\eps^3$ since $\rho_\eps=1+\mathcal{O}(\eps)$. Hence, bending does not contribute to the leading order of $H_\mathrm{a}$. Now inserting the explicit form of $\phi_0$ an elementary calculation yields
\begin{equation}
\label{eq:V_a hollow}
V_\mathrm{a} + \tilde V_{\mathrm{bend}}^\mathrm{a}  = 
\begin{aligned}[t]
 \int_{F_x}& - \tfrac{1}{2}\rho_\eps \Delta_\sH \norm{\rho_\eps}_1^{-1}
 - \tfrac{1}{2} g_B\bigl(\grad \abs{F_x}^{-1}, \pi_{M*} \grad \rho_\eps\bigr) \\
 &+ \tfrac{1}{4} \rho_\eps  \norm{\rho_\eps}_1 g_B\bigl(\pi_{M*} \grad \norm{\rho_\eps}_1^{-1}, \pi_{M*} \grad \norm{\rho_\eps}_1^{-1}\bigr)\ \dd g_F + \mathcal{O}(\eps^2)\, .
\end{aligned}
\end{equation}
With $\rho_\eps=1+\mathcal{O}(\eps)$ and $\Vert\rho_\eps\Vert_1= \vert F_x \vert + \mathcal{O}(\eps)$ one easily checks that up to order $\eps$ this expression equals
\[
 \tfrac{1}{4} g_B\bigl(\dd \log  \abs{F_x} , \dd \log \abs{F_x} \bigr) +
\tfrac{1}{2} \Delta_{g_B}\log \abs{F_x}  \, .
\]
As far as the remaining terms of $H_1$ are concerned, note that the scaled pullback metric $g^\eps$ of the hollow waveguide has the same expansion on horizontal one-forms up to errors of order $\eps^4$ as in the case of the massive waveguide~\eqref{eq:dual metric}, \ie
\[ g^\eps(\pi_M^*\dd x^i, \pi_M^*\dd x^j)=\eps^2 \bigl(g_B^{ij} + 2\eps \sff(\nu)^{ij} + \mathcal{O}(\eps^2)\bigr)\, .
\]
Hence, we can calculate these terms starting from expression~\eqref{eq:H_1 allg}. Since the latter all carry a prefactor $\eps^3$, we may replace any $\phi_0$ by $\abs{F_x}^{-1/2}$, obtaining for $\psi\in L^2(B)$
\begin{equation}
\label{eq:diff H_1 hollow}
\int_{F_x} \dive_{g_\mathrm{s}} \bigl(\abs{\phi_0}^2 \sff(\nu)(\dd\psi, \cdot)\bigr)\ \dd g_F =
\int_{F_x} \dive_{g_\mathrm{s}} \bigl(\abs{F_x}^{-1} \sff(\nu)(\dd\psi, \cdot)\bigr)\ \dd g_F + \mathcal{O}(\eps)
\end{equation}
and
\begin{align}
&\int_{F_x} \phi_0 \dive_{g_\mathrm{s}} \bigl(\sff(\nu)(\pi_{M*}\grad_{g_\mathrm{s}} \phi_0, \cdot)\bigr)\ \dd g_F\nonumber \\
&\ =\int_{F_x} \abs{F_x}^{-1/2} \dive_{g_\mathrm{s}}  \bigl(\sff(\nu)\bigl(\dd \abs{F_x}^{-1/2}, \cdot\bigr)\bigr)\ \dd g_F + \mathcal{O}(\eps)\, . \label{eq:pot H_1 hollow}
\end{align}
As in equation~\eqref{eq:diff H_1 massive} we have
\[
 \int_{F_x} \dive_{g_\mathrm{s}} \bigl(\abs{F_x}^{-1} \sff(\nu)(\dd\psi, \cdot)\bigr)\ \dd g_F
 = \dive_{g_B}\int_{F_x} \abs{F_x}^{-1} \sff(\nu)(\dd\psi, \cdot)\ \dd g_F\,.
\]
Again, this term vanishes if the barycentre of the fibres $F_x=\partial \mathring{F}_x \subset \sN_xB$ is zero, that is
\[
 \int_{F_x} \nu  \ \dd g_F=0\,.
\]
Since $\lambda_0\equiv 0$, the adiabatic operator is of the form
\begin{align}
H_\mathrm{a}&=-\eps^2 \Delta_{g_B} + \eps^2 V_\mathrm{a} + \eps P_0 H_1 P_0 \nonumber \\
&= -\eps^2 \Delta_{g_B} + \eps^2\Bigl( \tfrac{1}{4} g_B\bigl(\dd \log  \abs{F_x} , \dd \log \abs{F_x} \bigr) +
\tfrac{1}{2} \Delta_{g_B}\log \abs{F_x}\Bigr) + \mathcal{O}(\eps^3)\, , \label{eq:Vhollow}
\end{align}
with an error in $\mathcal{L}(W^2(B), L^2(B))$. Thus, the adiabatic operator at leading order is just the Laplacian on the base $B$ plus an effective potential depending solely on the relative change of the volume of the fibres. Going one order further in the approximation, we have
\[
\eps^2 V_\mathrm{a} + \eps P_0 H_1 P_0 =\eps^2\eqref{eq:V_a hollow} + 2\eps^3 \eqref{eq:pot H_1 hollow} + 2\eps^3\eqref{eq:diff H_1 hollow} + \mathcal{O}(\eps^4)
\]
in the same norm. Hence, $H_\mathrm{a}$ also contains the second order differential operator~\eqref{eq:diff H_1 hollow} if the barycentre of $F_x$ is different from zero. Let us also remark that the leading order of the adiabatic potential can also be calculated by applying a unitary transformation $L^2(F, \dd g_F)\to L^2(F, \abs{F_x}^{-1} \dd g_F)$ that rescales fibre volume to one, in the spirit of $\mathcal{M}_\rho$ (\cf equation~\eqref{eq:Vrhoeps}). In this way a similar potential was derived by Kleine~\cite{Kle}, in a slightly different context, for a special case with one-dimensional base and without bending.

\bibliographystyle{amsalpha}
 \providecommand{\bysame}{\leavevmode\hbox to3em{\hrulefill}\thinspace}   
 
\providecommand{\MR}{\relax\ifhmode\unskip\space\fi MR }
\providecommand{\MRhref}[2]{%
  \href{http://www.ams.org/mathscinet-getitem?mr=#1}{#2}
}
\providecommand{\href}[2]{#2}

\end{document}